\documentclass[authoryear, preprint,5p,times,twocolumn]{elsarticle}

\usepackage{amssymb}
\usepackage{multirow}
\usepackage{tabularx}
\usepackage{url}
\usepackage{amsmath}
\usepackage{xcolor}
\usepackage{graphicx} 
\usepackage{booktabs}
\usepackage[colorlinks,linkcolor=blue]{hyperref}

\begin{document}
\setlength{\parskip}{0pt}

\begin{frontmatter}

\title{FetalFlex: Anatomy-Guided Diffusion Model for Flexible Control on Fetal Ultrasound Image Synthesis}

\author{Yaofei Duan\fnref{label1}}

\author{Tao Tan\corref{cor1}\fnref{label1}}
\ead{taotan@mpu.edu.mo}

\author{Zhiyuan Zhu\fnref{label2,label7}}

\author{Yuhao Huang\fnref{label2,label7}}

\author{Yuanji Zhang\fnref{label2,label7}}

\author{Rui Gao\fnref{label3}}

\author{Patrick Cheong-Iao Pang\fnref{label1}}

\author{Xinru Gao\fnref{label9}}

\author{Guowei Tao\fnref{label8}}

\author{Xiang Cong\fnref{label8}}

\author{Zhou Li\fnref{label4,label5}}

\author{Lianying Liang\fnref{label4,label5}}

\author{Guangzhi He\fnref{label4,label5}}

\author{Linliang Yin\fnref{label6}}

\author{Xuedong Deng\corref{cor1}\fnref{label6}}
\ead{xuedongdeng@163.com}

\author{Xin Yang\corref{cor1}\fnref{label2,label7}}
\ead{xinyang@szu.edu.cn}

\author{Dong Ni\corref{cor1}\fnref{label2,label7}}
\ead{nidong@szu.edu.cn}

\cortext[cor1]{Corresponding author.}

\affiliation[label1]{organization={Faculty of Applied Sciences, Macao Polytechnic University},
            city={Macao},
            country={China}}

\affiliation[label2]{organization={National-Regional Key Technology Engineering Laboratory for Medical Ultrasound, School of Biomedical Engineering, Shenzhen University Medical School, Shenzhen University},
            city={Shenzhen},
            state={Guangdong},
            country={China}}

\affiliation[label3]{organization={Shenzhen RayShape Medical Technology Co., Ltd},
            city={Shenzhen},
            state={Guangdong},
            country={China}}

\affiliation[label4]{organization={Department of Ultrasound, Shenzhen Guangming District People’s Hospital},
            city={Shenzhen},
            state={Guangdong},
            country={China}}

\affiliation[label5]{organization={Jinan University},
            city={Guangzhou},
            state={Guangdong},
            country={China}}
            
\affiliation[label6]{organization={Center for Medical Ultrasound, The Affiliated Suzhou Hospital of Nanjing Medical University, Suzhou Municipal Hospital, Gusu School, Nanjing Medical University},
            city={Suzhou},
            state={Jiangsu},
            country={China}}

\affiliation[label7]{organization={Medical Ultrasound Image Computing (MUSIC) Laboratory, Shenzhen University},
            city={Shenzhen},
            state={Guangdong},
            country={China}}

\affiliation[label8]{organization={Qilu Hospital of Shandong 
 University},
            city={Jinan},
            state={Shandong},
            country={China}}

\affiliation[label9]{organization={Northwest Women \& Children Hospital},
            city={Xian},
            state={Shaanxi},
            country={China}}

\begin{abstract}
Fetal ultrasound (US) examinations require the acquisition of multiple planes, each providing unique diagnostic information to evaluate fetal development and screening for congenital anomalies. 
However, obtaining a comprehensive, multi-plane annotated fetal US dataset remains challenging, particularly for rare or complex anomalies owing to their low incidence and numerous subtypes. 
This poses difficulties in training novice radiologists and developing robust AI models, especially for detecting abnormal fetuses.
In this study, we introduce a Flexible Fetal US image generation framework (FetalFlex) to address these challenges, which leverages anatomical structures and multimodal information to enable controllable synthesis of fetal US images across diverse planes. 
Specifically, FetalFlex incorporates a pre-alignment module to enhance controllability and introduces a repaint strategy to ensure consistent texture and appearance. 
Moreover, a two-stage adaptive sampling strategy is developed to progressively refine image quality from coarse to fine levels. 
We believe that FetalFlex is the first method capable of generating both in-distribution normal and out-of-distribution abnormal fetal US images, without requiring any abnormal data. 
Experiments on multi-center datasets demonstrate that FetalFlex achieved state-of-the-art performance across multiple image quality metrics. 
A reader study further confirms the close alignment of the generated results with expert visual assessments. 
Furthermore, synthetic images by FetalFlex significantly improve the performance of six typical deep models in downstream classification and anomaly detection tasks.
Lastly, FetalFlex's anatomy-level controllable generation offers a unique advantage for anomaly simulation and creating paired or counterfactual data at the pixel level. The demo is available at: \href{https://dyf1023.github.io/FetalFlex/}{https://dyf1023.github.io/FetalFlex/}.
\end{abstract}


\end{frontmatter}

\section{Introduction}
\label{intro}
In fetal screening, radiologists should adjust the positions and angles of the ultrasound (US) probe to obtain multiple fetal planes.
These planes are vital for assessing fetal growth, monitoring pregnancy, and identifying congenital abnormalities~\citep{1-1}. 
However, frequent fetal movements and the presence of small or similar anatomical structures result in a time-consuming and labor-intensive process of acquiring high-quality US images~\citep{1-6}.
Moreover, abnormal samples in clinical practice are typically scarce and of many subtypes.
Hence, collecting abundant balanced and annotated images to train deep learning-based models for fetal US analysis remains difficult~\citep{2-12}.
Several studies \citep{dav,ven} have employed anomaly detection models trained exclusively on normal data to address the performance degradation typically caused by the scarcity of abnormal data.
However, most of these models cannot perform accurate fine-grained classification for abnormalities. 
The alternative approach is adopting traditional data augmentation techniques, e.g., rotation, flipping, and brightness adjustment, to expand the samples.
However, despite being easy to implement, their effectiveness in controllability and enhancing dataset diversity remains limited~\citep{frid2018gan,gui2021review}. 

\begin{figure}[t]
    \centering
    \includegraphics[width=1\linewidth]{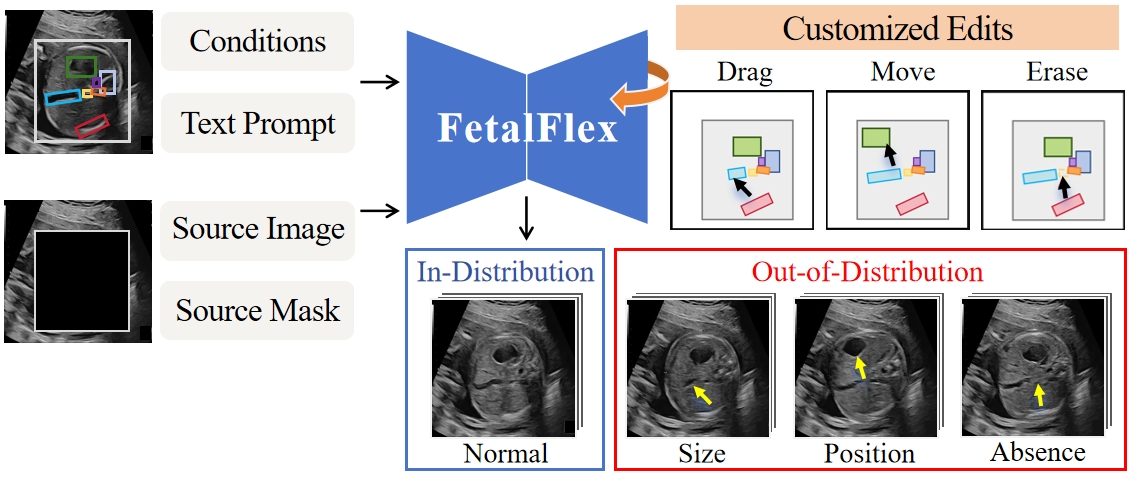}
    \caption{Overview of the proposed framework. The framework is designed to enabling user-editable generation, catering to both in-distribution (ID) and out-of-distribution (OOD) scenarios.}
    \label{fig0}
\end{figure}

Recently, generative models, i.e., generative adversarial networks (GANs)~\citep{gan}, variational autoencoders (VAEs)~\citep{vae}, and diffusion models~\citep{diffusion}, have demonstrated the potential to produce realistic synthetic data, offering promising solutions to the abovementioned challenges. 
Diffusion models have recently gained significant attention owing to their flexible conditional architecture compared to that of competitors.
However, the value of diffusion-driven medical image synthesis remains under-explored, and several key challenges remain to be addressed. 
\textit{First}, general diffusion model architectures and sampling strategies tailored specifically for fetal US images involving multiple planes are currently lacking. 
\textit{Second}, the scarcity of medical data and annotations restricts the ability of diffusion models to capture complex characteristics, particularly for abnormal samples of limited quantity but significant medical value.
Guiding models to generate abnormalities in scenarios with few or no available abnormal images and absent reference knowledge of abnormal categories remains an unsolved problem.
\textit{Third}, the generated results must meet the clinician's visual assessment criteria and ensure image-level interpretability to enhance their confidence \citep{heal2}, while also effectively supporting various downstream tasks for different models.
\textit{Last}, fulfilling the user’s requirements for controllable and editable US images during the generation is a key advantage for practical applications. Our objective is to develop a framework, as illustrated in Fig. \ref{fig0}, that addresses the aforementioned requirements to generate valuable fetal US images for both ID and OOD scenarios.

Existing studies on medical image synthesis have largely overlooked the role of text-based control in guiding the synthesis of specific categories during the generation process~\citep{shenda,5500}. 
Previous models were often limited to synthesizing a single type of medical image, neglecting shared characteristics across different categories. Therefore, these models typically require re-training from scratch to address new synthesis tasks.
More critically, they rely on extensive training with large datasets of medical images. For example, researchers trained models on 60K images for pharyngeal synthesis \citep{1-44} and on 25K slices for accelerated MRI reconstruction~\citep{D3}. However, acquiring such large-scale annotated datasets in the fetal US is challenging.
Many studies introduced control conditions during the generation process to enhance the quality of generated images \citet{qiao,1-56}.
For instance, \citet{shenda} introduced edge sketches and target masks as control conditions.
However, they mainly focused on cases with simple anatomical structures, such as the lungs, hips, and ovaries. 
While~\citet{1-11} guided the fetal heart synthesis using canny edges and segmentation maps, the model lacked the customization options for image editing.
Most studies focused only on generating adequate normal samples; however, encountering unusual cases is common in clinical practice, underscoring the urgent need to synthesize abnormal cases~\citep{1-12}.
Certain studies~\citep{2-11,1-55} have aimed to generate abnormal cases; however, their models typically depended on learning from large quantities of positive data to shift the data distribution of synthesized positive samples toward the decision boundary at the feature-level.
However, abnormal cases are diverse in fetal US, and the proportion of abnormal samples can be as low as 1 in 1K or even 1 in 10K~\citep{21}; therefore, relying on large datasets of positive samples is impractical in fetal US.
Moreover, previous studies focusing on altering data distributions at the high-dimensional feature-level often resulted in generated anomalies lacking interpretability at the image-level, thus failing to meet the visual requirements.
Due to the high variability of fetal anatomical structures, fetal US imaging requires higher visual standards, making the development of interpretable generative models particularly challenging.
This requires generative models that capture fine anatomical details while providing visual cues that meet expert knowledge and diagnostic criteria.
Therefore, we propose FetalFlex, a universal framework for fetal US image generation. 
To the best of our knowledge, this is the first method capable of generating multiple planes, offering controlled generation, diverse outputs, and allowing user-editable synthesis of both normal and abnormal fetal US images, without compromising image fidelity. 
Our proposed method can generate various distinct types of fetal US planes, including the thalamic transverse plane, facial sagittal plane, and upper abdominal transverse plane, using a single universal model, thus eliminating the need for re-training. 

\begin{figure*}[!h]
    \centering
    \includegraphics[width=1\linewidth]{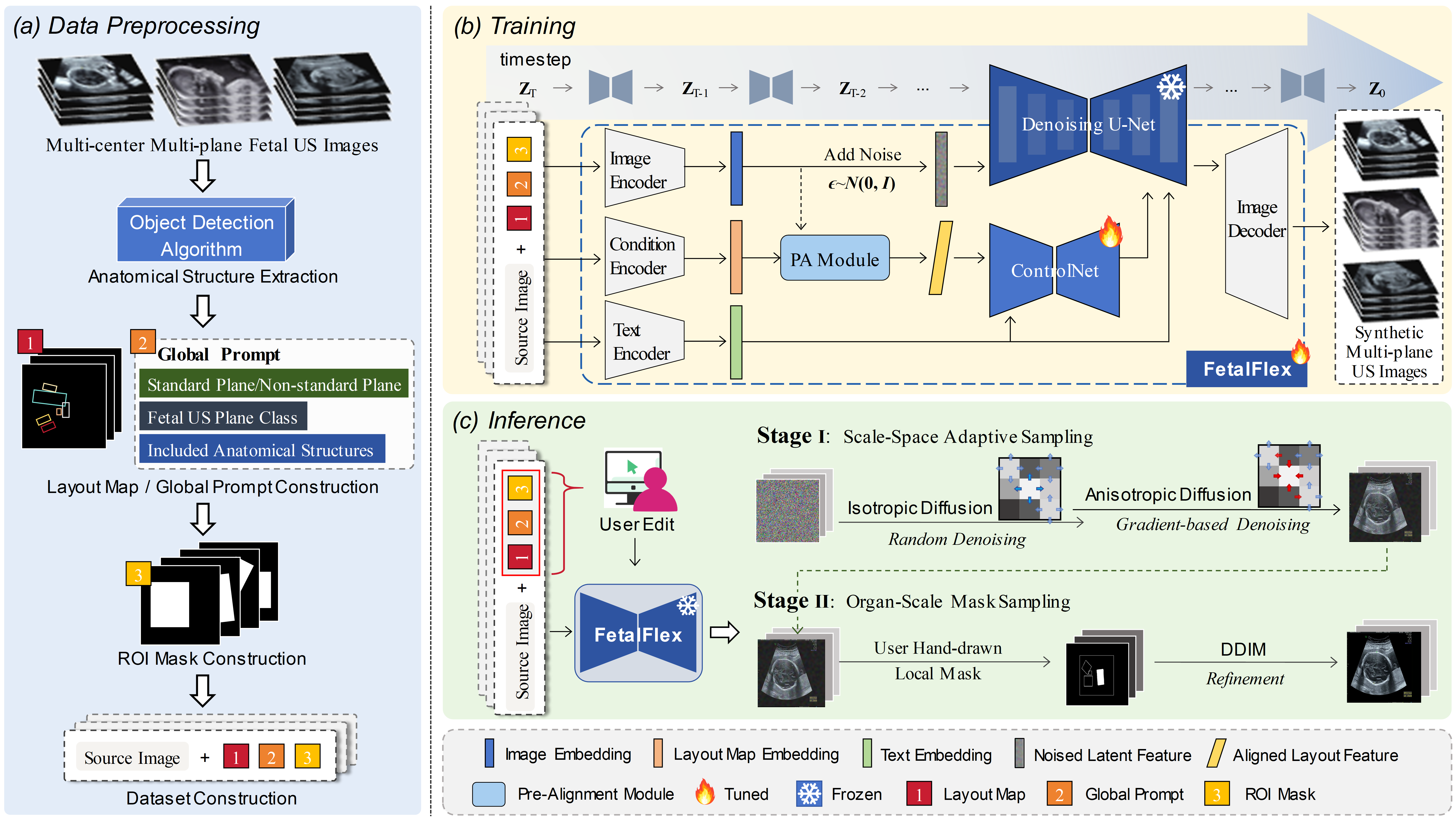}
    \caption{Overview of the proposed FetalFlex framework: (a) Data Preprocessing. The anatomical structures of fetal US images are extracted using detection algorithms FTSPD \citet{zhen}. Layout maps, global prompts, and ROI masks are subsequently constructed from the bounding box information as control conditions. (b) Training. During the training process, the model is finetuned on the known US background according to the provided control conditions. (c) Inference. During the sampling process, users could edit the control conditions, enabling the generation of controlled fetal US images across multiple planes through a two-stage sampling procedure.}
    \label{wf}
\end{figure*}

Our contributions can be summarized as follows:

\begin{itemize}
    \item We propose a novel and general fetal US image generation framework named FetalFlex. 
    It uses one unified model to support flexible and controllable generation for multiple fetal planes without re-training.
    \item We introduce a rough mask to localize the region of interest (ROI), ensuring that FetalFlex can generate appearance-consistent US images on a small dataset. 
    Additionally, we propose a two-stage sampling strategy based on US imaging characteristics, i.e., grayscale differences.
    \item We incorporate a pre-alignment module to enhance the controllability of anatomical information for generating both normal and abnormal samples. FetalFlex can simulate anomalies such as variations in anatomical structure size, structural displacement, or structural deficiencies.
    \item Extensive experiments on multi-center datasets demonstrate that FetalFlex achieves SOTA performance in image quality and aligns with radiologists' visual preferences. 
    Two downstream tasks additionally verify that the generated results significantly improved the performance of various deep models.
\end{itemize}

\section{Related Work}
\label{Related Work}
\textbf{Deep Learning in Fetal US.}
Deep learning-based methods have potential applications in obstetric US~\citep{2-12,2-15,2-16}, including measuring fetal biometric parameters and identifying normal and abnormal fetal anatomical structures, etc. 
Several studies \citep{harikumar,krishna2023,krishna,huang2023fourier} have developed deep learning models to automatically classify common fetal US planes, aiming to improve detection efficiency and accuracy. Additionally, studies \citep{komatsu2021detection, xie2020using} have proposed supervised frameworks for detecting abnormalities in the cardiac and brain of the fetus, aiming to enhance the sensitivity of basic clinical examinations.
Previous intelligent methods have achieved significant advancements in normal fetal analysis; however, the overall detection rate of anomalies remains relatively low~\citep{2-17}.
This challenge primarily arises from the diverse and complex nature of fetal anomalies, complicating the task for deep models to recognize a uniform feature set.
The rarity of abnormalities and the differences in incidence rates further exacerbate the shortage and imbalance of positive samples for model training~\citep{2-14}.
Moreover, collecting large-scale annotated datasets is extremely difficult owing to ethical and privacy concerns. Synthetic data presents a promising solution by supplying ample training samples without the need to share real patient data, thereby reducing the risk of data leakage.

\textbf{Medical Image Synthesis.} 
Generative models have gained significant attention in image generation in recent years.
In the earliest studies, GANs have been widely used to synthesize CT, MRI, and X-ray images~\citep{2-1,2-2,2-3}. 
However, GANs encounter challenges in synthesizing complex multi-modal distributions~\citep{high}. 
Distinct from GANs, diffusion models can generate realistic samples with complex patterns by progressively denoising images from noise, offering a more stable training process and greater output diversity.
\citet{D3} proposed score-based diffusion models for accelerated MRI reconstruction. 
Another study \citep{2-7} adopted diffusion models to synthesize 2D chest X-rays, cardiac MRIs, and CT scans, achieving resolutions of 256×256. 
CoLa-Diff \citep{1-56} further improved brain MRI synthesis by introducing brain region masks as priors in the diffusion process.
However, the aforementioned studies focused solely on normal images and overlooked the more critical synthesis of abnormal samples in the medical image field.

\begin{figure*}[!h]
    \centering
    \includegraphics[width=0.96\linewidth]{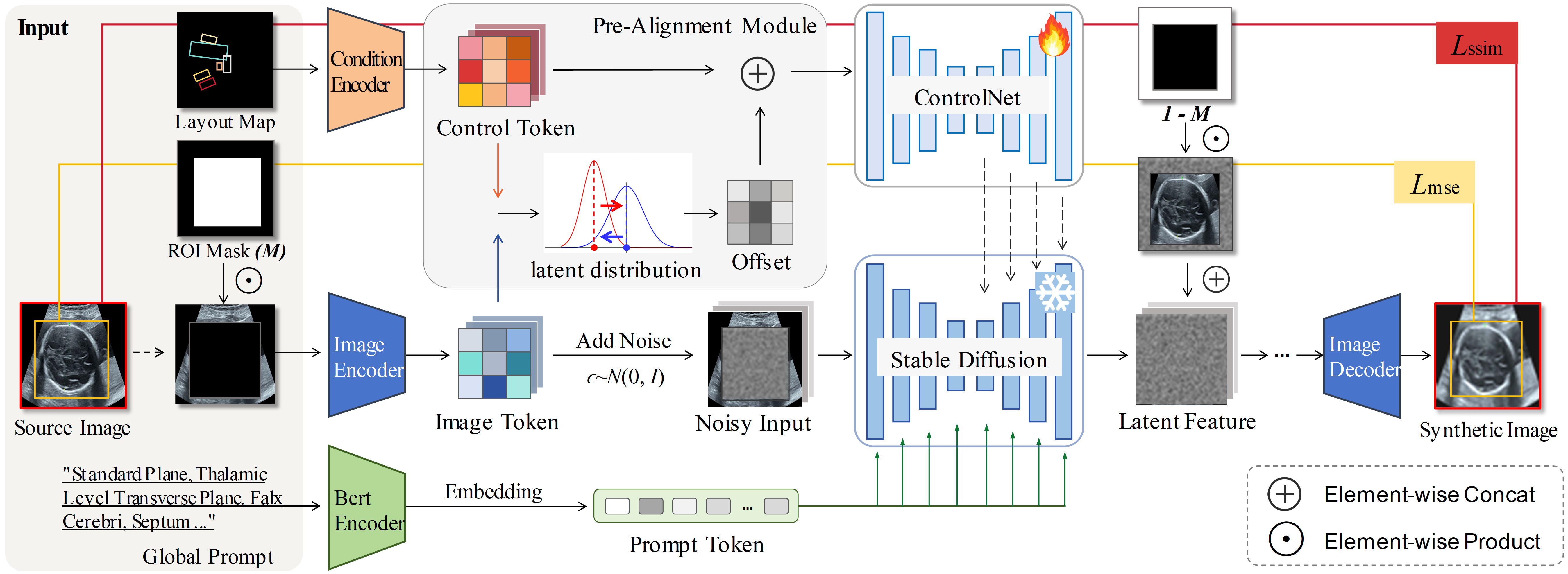}
    \caption{Schematic of the proposed FetalFlex. The model processes multimodal control conditions through a combination of encoders, a pre-alignment module, a denoising U-Net, and ControlNet, to achieve controllability.}
    \label{network}
\end{figure*}

\textbf{Abnormal Medical Image Synthetic.}  
The synthesis of abnormal images has not received widespread attention, despite being fundamental to anomaly detection \citep{heal3}.
Deep learning-based diagnosis systems often struggle to obtain a sufficient number of manually labeled data for training owing to the diversity, complexity, and rarity of abnormal samples, thereby limiting their generalizability to practical applications. 
A previous study \citep{2-10} used GAN to synthesize liver lesion images to assist in training classification models, whereas others \citep{2-11} applied diffusion model-based sampling strategies to guide the generation of samples near class boundaries for fundus, breast, lung, and skin images.
~\citet{1-55} extended the repaint method to simulate stroke-induced brain MRI lesions in mask regions. 
However, these methods focused on synthesizing positive samples with pathological changes and relied on learning image features from a large number of positive samples. 
Unlike synthesizing pathological features, abnormal fetal US imaging involves significant morphological and structural variations. 
Thus, models for synthesizing abnormal fetal US images must specifically generate samples with varied anatomical structures.

In summary, existing medical image synthesis models often fall short in the fetal US due to the limited scalability of model architectures across different categories (e.g., handling multiple fetal planes using one model is challenging) and the extensive data required for fine-tuning. 
Therefore, in this study, we aimed to design a universal framework for synthesizing fetal US planes to enable editable synthesis of both normal and abnormal samples are our key focuses.

\section{Methodology}
\label{method}
Fig.~\ref{wf} shows the workflow of the FetalFlex framework.
This framework obtains multiple anatomical structures and corresponding information from various US planes as input and generates corresponding fetal US images in a flexible and controllable manner. 
The following sections will provide additional technical details. 
Section \ref{3.1} introduces the preliminaries of diffusion models, and Section \ref{3.3} illustrates the implementation details, including key design modules and the two-stage anatomy-to-image generation process.

\subsection{Preliminaries}
\label{3.1}
We first utilize a pre-trained Latent Diffusion Model (LDM) with ControlNet as the backbone of our framework. 
LDMs involve a T-step forward diffusion process and a T-step reverse denoising process from a prior distribution. In essence, the forward process incrementally introduces random Gaussian noise of varying intensities to a clean data point \textit{z0}, whereas the reverse process aims to gradually remove the noise to reconstruct \textit{z0} by a generative \textit{Markov} chain.
The forward-time diffusion process can be represented as:

\begin{equation}
    x_{t}=\sqrt{\bar{\alpha}_{t}} x_{0}+\sqrt{1-\bar{\alpha}_{t}} \epsilon,
\end{equation}
where $\sqrt{\bar{\alpha}_{t}}$ denotes the level of noise added at timestep t.
Besides, $\epsilon$ is noise sampled from a standard normal distribution \textit{N(0, I)} with mean=0 and variance=1.

Gaussian noise is reversed back into samples from the learned distribution by the diffusion model to generate new images. This process can be modeled as the following equation with a neural network $\epsilon_{\theta}$ for predicting noise:

\begin{equation}
    \mathcal{L}_{D M}=\mathbb{E}_{x, \epsilon \sim \mathcal{N}(0, \mathbf{I}), t}\left[\left\|\epsilon-\epsilon_{\theta}\left(x_{t}, t\right)\right\|_{2}^{2}\right].
\end{equation}

Denoising Diffusion Implicit Models (DDIM) break the limitations of the \textit{Markov} chain \citep{3-0} and further achieve accelerated sampling, without altering the forward noise process:

\begin{equation}
    x_{t-1}=\sqrt{\bar{\alpha}_{t-1}} \hat{x}_{0 \mid t}+\sqrt{1-\bar{\alpha}_{t-1}-\sigma_{t}^{2}} \epsilon_{\theta}\left(x_{t}, t\right)+\sigma_{t} \epsilon,
\end{equation}
where $\epsilon_{\theta}\left(x_{t}, t\right)$ is a learned approximation of the noise $\epsilon$ that corrupted the original image $x_{0}$ to produce $x_{t}$, which can be parameterized with a U-Net architecture.

Furthermore, ControlNet~\citep{controlnet} enhances the generation controllability by effectively and efficiently replicating parts of the SD's architecture and weights, allowing it to learn various control conditions. The corresponding objective can be simplified as follows:

\begin{equation}
    \left.\mathcal{L}=\mathbb{E}_{\boldsymbol{z}_{0}, \boldsymbol{t}, \boldsymbol{c}_{t}, \boldsymbol{c}_{f}, \epsilon \sim \mathcal{N}(0,1)}\left[\| \epsilon-\epsilon_{\theta}\left(\boldsymbol{z}_{t}, \boldsymbol{t}, \boldsymbol{c}_{t}, \boldsymbol{c}_{\mathrm{f}}\right)\right) \|_{2}^{2}\right],
\end{equation}
where ${z}_{0}$ represents input image, $t$ represent timestep.
${c}_{t}$ and ${c}_{f}$ denote text prompts and task-special conditions, respectively.
The model trains the network $\epsilon_{\theta}$ to predict the noisy image ${z}_{t}$.

\subsection{FetalFlex for Universal Fetal US Image Generation}
\label{3.3}
Fig.~\ref{network} illustrates the architecture of the FetalFlex. 
It first embeds multimodal control conditions into their respective encoders to obtain multimodal tokens. The control tokens are subsequently aligned with the image tokens along the feature dimension and fed into the ControlNet branch. Meanwhile, the prompt tokens and timestep information, along with the noise-added image tokens, are input into the SD network to learn the denoising process in reverse. We employ the Repaint strategy, where the ROI mask is applied to the intermediate results to refine the output. Additionally, we design a sampling strategy tailored to US data that transitions from coarse to fine granularity, enabling realistic and diverse US image synthesis. In this section, we will provide a detailed description of the model architecture (see Section \ref{3.3.1}), discuss the style-consistent training strategy in Section \ref{3.3.2}, and introduce the two-stage sampling strategy in Section \ref{3.3.3}. Finally, we introduced the loss function we employed \ref{3.3.4}.

\subsubsection{Model Architecture}
\label{3.3.1}
We aimed to synthesize fetal US images that are globally consistent in appearance, and locally diverse in ROI areas.
Moreover, we hope the generation should be user-controllable and that the generated results are realistic.

Hence, we carefully design the model architecture to achieve the abovementioned objectives. 
We use a pre-trained VAE as both the image and condition encoders, mapping images and layout maps (512x512) into latent features (64x64). 
Meanwhile, we replace CLIP with a BERT pretrained on a large medical dataset of PubMed abstracts as the text encoder \citet{3-1}, extracting the last hidden state of BERT as the prompt embedding features. 
Since CLIP is primarily trained on open-domain visual and textual data, it may lack a specialized understanding of medical data. 
This limits FetalFlex from becoming a general model for generating multi-plane images controlled by textual input.
In contrast, the pre-trained BERT captures rich medical semantic information from the global prompt, effectively connecting the semantics of US planes and anatomical structures to corresponding image features. 

Additionally, we innovatively adopt layout maps as one of the control conditions. 
However, a significant disparity existed in data distribution between features of the layout maps and those of the images owing to the high sparsity of layout maps after embedding.
Thus, we introduce the image-layout map pre-alignment module to better guide the model’s attention to the relationship between the layout maps and images and precisely direct the anatomical structure's positioning.
Specifically, we rescale and shift the feature space of the layout map at the pixel-level to align it with the image, subsequently merging it channel-wise back into the layout map. 

During training, we freeze the weights of the denoising U-Net and load pre-trained ControlNet weights from natural images for fine-tuning. Additionally, we introduced a 30\% probability of replacing the layout map and global prompt with null values to enhance the model's ability to recognize the semantics of the control conditions.

\subsubsection{Conditional Inpainting}
\label{3.3.2}
Fine-tuning a large-scale SD pre-trained on millions of natural images using a small fetal US dataset may yield suboptimal results~\citep{xie2023difffit}. 
This is primarily attributed to the significant differences in data distribution between medical and natural images. 
To address this issue,~\citet{3-01} first introduced an inpainting method (named RePaint), which constrains the denoising process within masked areas by using unmasked regions as contextual references during denoising. 

However, the model tends to reproduce the learned data distribution when handling large-masked regions, resulting in synthetic results that often appear random.
Inspired by RePaint, we optimize it to create a novel inpainting pipeline. 
It extracts the ROI from fetal US images as the binary masked region, leveraging global prompt information and fetal US-specific anatomical structures as control conditions, while ensuring content consistency in appearance between the ROI and the background during the generation process.

In each reverse step $i \in\{T, T-1, \ldots, 2\}$, we achieve the following expressions:
\begin{equation}
    \begin{aligned}
        x_{t-1}^{bg} & \sim N\left(\sqrt{\bar{a}_{t}} x_{0}^{bg},\left(1-\bar{a}_{t}\right) I\right), \\
        x_{t-1}^{RoI} & \sim N\left(\epsilon_{\theta}\left(x_{t}, x_{\text {mask }}, c_{t}, c_{t}, t\right), \Sigma_{\theta}\left(x_{t}, t\right)\right), \\
x_{t-1} & =m \odot x_{t-1}^{bg}+(1-m) \odot x_{t-1}^{RoI} \\
& =m \odot x_{t-1}^{bg}+\left(1 - m) \odot \text { DDIM } \left(\epsilon _{\theta}\left(x_{t}, x_{\text {mask }},c_{t}, c_{f}, t\right)\right)\right..
\end{aligned}
\end{equation}
Thus, $x_{t-1}^{bg}$ is sampled using the given pixels in the original image $m \odot x$, whereas $x_{t-1}^{RoI}$ is sampled from the model.
In the final timestep, we employ an unmasked DDIM step to ensure the coherence between the denoised features in the ROI and the background regions. The denoised features are then fed into the decoder to generate the final US image.

\subsubsection{Two-Stage Sampling Strategy}
\label{3.3.3}
The two-stage sampling strategy is designed to overcome the limitations of single-stage generation methods~\citep{3-4-1,3-4-2,3-4-3}. 
In this study, we introduced a coarse-to-fine random-to-specified multi-scale approach during the inference stage of the diffusion model. 
A detailed explanation is as follows:

\textbf{Stage 1. Scale-Space Adaptive (SSA) Sampling.} Pixel intensity distribution in medical US images often deviates from the Gaussian model owing to the nature of US imaging, which constructs tissue images by reflecting sound waves. This deviation poses a challenge for diffusion models that assume independent pixel values are independent and follow a Gaussian distribution. In the principle of US imaging, there exists a phenomenon known as  ``acoustic impedance difference", where different tissues have distinct acoustic properties, resulting in pronounced grayscale differences at anatomical structure edges in the fetal US images.
Based on the abovementioned characteristics, we propose SSA sampling, an adaptive denoising method operating across different scale spaces based on the content of US images. 
Specifically, we incorporate the concept of Anisotropic Diffusion (AD)~\citep{PM} into the sampling process. 
AD, grounded in partial differential equations (PDEs), calculates the gradient of grayscale variations around each pixel to determine whether the pixel is at an organ boundary, thereby dynamically adjusting the diffusion's intensity and direction, as shown in the equation \ref{ee6} and equation \ref{ee7}. Unlike DDIM's isotropic diffusion (ID) approach, this method focuses on enhancing tissue boundaries, rather than allocating equal attention to all anatomical structural information.

\begin{equation}
    A_{n i} x_{t}=\nabla \cdot\left(c\left(\left\|\nabla x_{t-1}\right\|\right) \cdot \nabla x_{t-1}\right),
    \label{ee6}
\end{equation}

\begin{equation}
    \begin{array}{l}
c\left(\left\|\nabla x_{t-1}\right\|\right)=\exp \left(-\left(\left\|\nabla x_{t-1}\right\| / k\right)^{2}\right),
\label{ee7}
\end{array}
\end{equation}
\\
where $\nabla$ represents divergence operator, $c\left(\left\|\nabla x_{t-1}\right\|\right)$ represents diffusion coefficient, and $\left\|\nabla x_{t-1}\right\| $ means pixel gradient.

In AD sampling, the gradient of each pixel is calculated and the image is denoised using an iterative approach: 
\begin{equation}
    A_{n_{i}} x_{t+1}=A_{n i}x_{t}+\delta \cdot \sum_{i}^{i \in\{U, D, L, R\}}\left(C_{i} \cdot \nabla_{i} A_{n i} x_{t}\right),
    \label{e4}
\end{equation}
\\
where $\delta$ represents the magnitude of the update for each timestep, and ${i \in\{U, D, L, R\}}$ means calculate the gradient of the pixel in four directions (Up, Down, Left, Right), where the gradient is typically large at the edge of the organ.

SSA sampling applies an adaptive AD algorithm at each timestep to the intermediate values generated in the DDIM step. The SSA sampling can be expressed as follows:

\begin{equation}
x_{t}=\left(A_{ni}x_{t-1}\ominus Ix_{t-1}\right)\oplus Ix_{t-1},
\end{equation}
\\
where $\ominus$ and $\oplus$ denote pixel-wise subtraction and pixel-wise addition, respectively.

\textbf{Stage 2. Organ-Scale Mask (OSM) Sampling.}
After Stage 1, we obtain a coarse-grained target image. 
We propose a fine-grained editable OSM sampling strategy to further enhance the quality and texture details of synthesized abnormal images and allow for interactive editing of synthetic samples during the generation process. 
OSM sampling takes the draft image (generated in the SSA stage or from other models) as input and uses a user-drawn mask (typically for constructing abnormal organ regions) as guidance for secondary optimization, allowing for precise control over local anatomical changes to generate abnormalities. 
Moreover, OSM sampling can serve as a post-processing tool to improve generated image quality.

\subsubsection{Ensemble Learning}
\label{3.3.4}
We designed distinct loss functions to handle the ROI and global image in the FetalFlex model. For ROI, which contains rich diagnostic information, we applied the Mean Squared Error (MSE) loss to ensure pixel-level precision. For the global image, we incorporated the Structural Similarity Index Measure (SSIM) loss to better capture the visual characteristics required for medical imaging. We aim to enhance both local and global image fidelity by ensembling local and global losses, balancing precision and visual coherence in US image synthesis. The ensemble loss function is defined as follows:
\begin{equation}
L_{\text {total }}=\lambda_{1} L_{MSE} * \text { mask }[:, \text { None, } :, :]+\lambda_{2} L_{S S I M},
\end{equation}
where $\lambda_{1}$ and $\lambda_{2}$ are the tuning parameters used to balance the contributions of the local MSE and the global SSIM.

\section{Experimental setting}
\subsection{Datasets}
We retrospectively collected fetal US images from seven medical centers that comprise our in-house balanced dataset, including three planes of fetal US images: thalamic transverse planes, facial sagittal planes, and upper abdominal transverse planes. 
This study was approved by the local institutional review boards.
We obtained 3,600 fetal US images from women between 20–24 (+6) weeks of gestation across multiple centers. 
The radiologists excluded images of unsatisfactory quality; nevertheless, the dataset still included both standard planes and non-standard planes under quality control by experienced radiologists.
Standard planes are defined according to the ISUOG Practice Guidelines~\citep{ISUO} and Practice Guidelines for Performance of Prenatal Ultrasound Screening (PPUS)~\footnote{\url{10.3760/cma.j.cn131148-20211110-00821}}, encompassing key anatomical views required for clinical diagnosis. 
Non-standard planes refer to images that do not fully meet the strict criteria of the guidelines, but still display clear anatomical structures and possess some diagnostic value. These non-standard planes were included in the dataset to enhance the model's generalization capability.

\begin{figure}[t]
    \centering
    \includegraphics[width=0.88\linewidth]{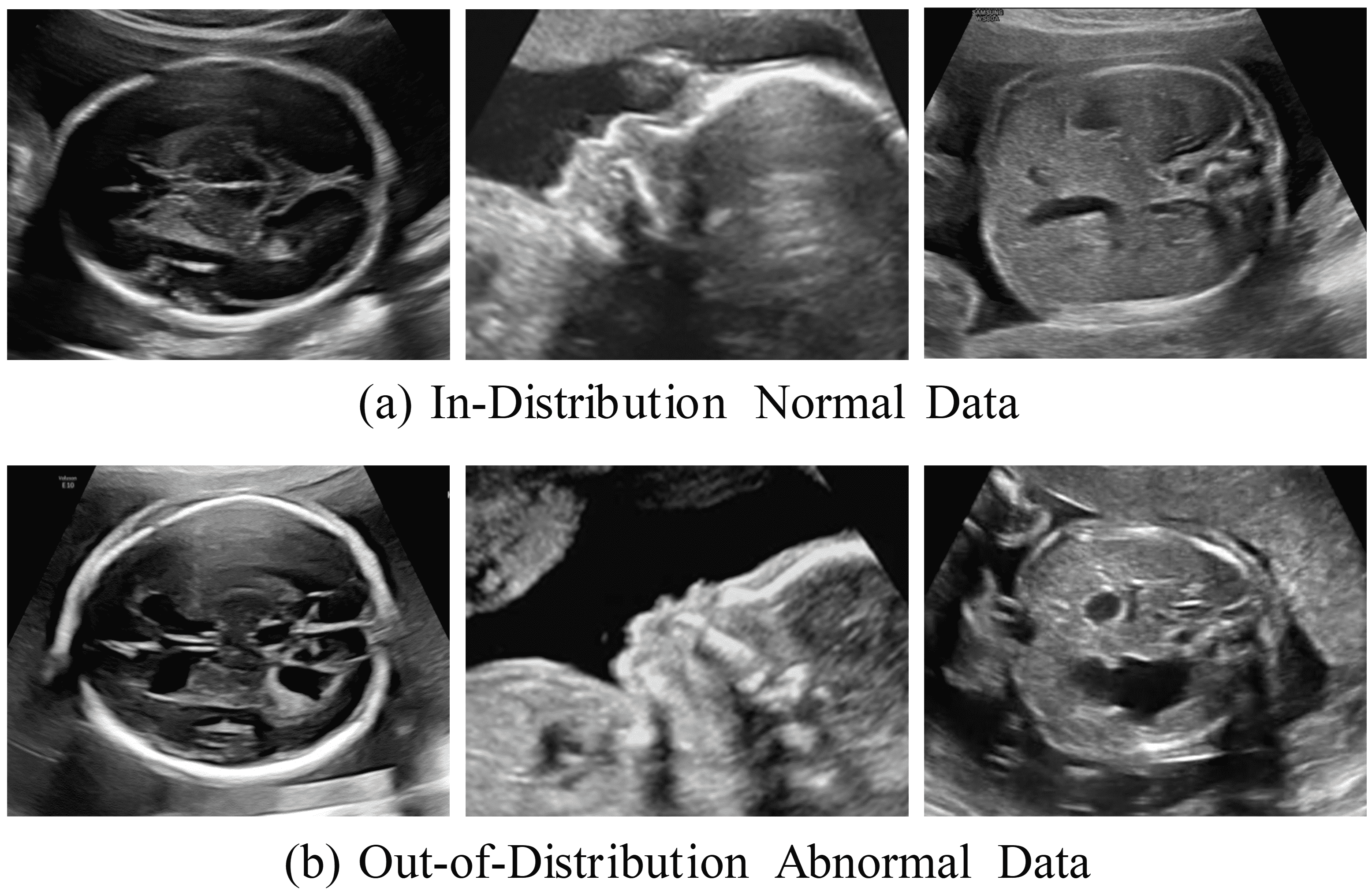}\\ 
    \caption{Examples of ID samples and corresponding OOD anomalies across three fetal planes.}
    \label{dataset}
\end{figure}

We randomly selected 800 samples for each plane as the training set, with the remaining 400 samples were used as the test set. 
We employed the well-trained FTSPD model (average accuracy=85\%)~\citep{zhen} to automatically generate bounding boxes for anatomical structures in each image to further reduce the annotation workload.
These bounding boxes were subsequently reviewed by radiologists to ensure precision and transformed into layout maps and ROI masks, with the corresponding category labels sorted to form text prompts.

Moreover, we retrospectively collected 360 real abnormal samples of significant fetal malformations to construct the anomaly detection experiments for downstream tasks, with 120 abnormalities for each type of US plane.
In detail, we collected real cases of fetal hydrocephalus in the thalamic dataset, samples of cleft lip and palate in the facial sagittal planes, and samples of duodenal atresia or stenosis in the upper abdominal transverse planes. 
Fig.~\ref{dataset} depicts typical normal samples in the dataset and abnormal US images with the aforementioned congenital malformations. 
These abnormal samples were diagnosed by experienced senior fetal US specialists according to the ISUOG and PPUS guidelines mentioned above.

\begin{table*}[t]
\centering
\caption{Comparisons with other methods in diverse fetal US image synthesis through four image quality metrics: PSNR, MS-SSIM, FID, and LPIPS. The ControlNet method is the baseline method, with blue font representing a decrease compared to the baseline value, red and dark gray font representing an increase.}
\resizebox{1\textwidth}{!}{
\begin{tabular}{l|cccc|llll}
\toprule
\multirow{2}{*}{Methods} & \multicolumn{4}{c|}{Conditions}      & \multicolumn{4}{c}{Metrics}                                                                                            \\
                         & Text & Canny & Layout map & ROI mask & \multicolumn{1}{c}{PSNR↑} & \multicolumn{1}{c}{MS-SSIM↑} & \multicolumn{1}{c}{FID↓} & \multicolumn{1}{c}{LPIPS(Alex)↓} \\ \midrule
Stable Diffusion v1.5~\citep{diffusion}        & \checkmark    &   -    &       -     &     -     & 8.874 \footnotesize{\textcolor{blue!85!black}{(-0.856)}}          & 0.098 \footnotesize{\textcolor{blue!85!black}{(-0.184)}}         & 171.952 \footnotesize{\textcolor{blue!85!black}{(-17.793)}}        & 0.675 \footnotesize{\textcolor{blue!85!black}{(-0.107)}}          \\
ControlNet \citep{controlnet}              &   -   &   -    & \checkmark          &     -     & 9.730 \footnotesize{\textcolor{gray}{(+0.0)}}           & 0.282 \footnotesize{\textcolor{gray}{(+0.0)}}         & 154.159 \footnotesize{\textcolor{gray}{(+0.0)}}       & 0.568 \footnotesize{\textcolor{gray}{(+0.0)}}         \\
Uni-ControlNet \citep{uni}          &   -   & \checkmark     & \checkmark          &     -     & 10.343 \footnotesize{\textcolor{black}{(+0.613)}}         & 0.358  \footnotesize{\textcolor{black}{(+0.076)}}        & 130.471 \footnotesize{\textcolor{black}{(+23.688)}}        & 0.518 \footnotesize{\textcolor{black}{(+0.050)}}         \\
Repaint \citep{3-01}                &   -   &    -   &   -         & \checkmark        & 20.266 \footnotesize{\textcolor{black}{(+10.536)}}        & 0.755 \footnotesize{\textcolor{black}{(+0.473)}}         & 59.254 \footnotesize{\textcolor{black}{(+94.905)}}          & 0.175 \footnotesize{\textcolor{black}{(+0.393)}}         \\
\textbf{FetalFlex (Ours)} & \checkmark    &   -    & \checkmark          & \checkmark        & \textbf{23.686} \footnotesize{\textcolor{red!90!black}{(+13.956)}} & \textbf{0.876} \footnotesize{\textcolor{red!90!black}{(+0.594)}} & \textbf{51.103} \footnotesize{\textcolor{red!90!black}{(+103.056)}} & \textbf{0.144} \footnotesize{\textcolor{red!90!black}{(+0.424)}} \\ \bottomrule
\end{tabular}
}
\label{c1}
\end{table*}

\begin{figure*}[!h]
    \centering
    \includegraphics[width=0.98\linewidth]{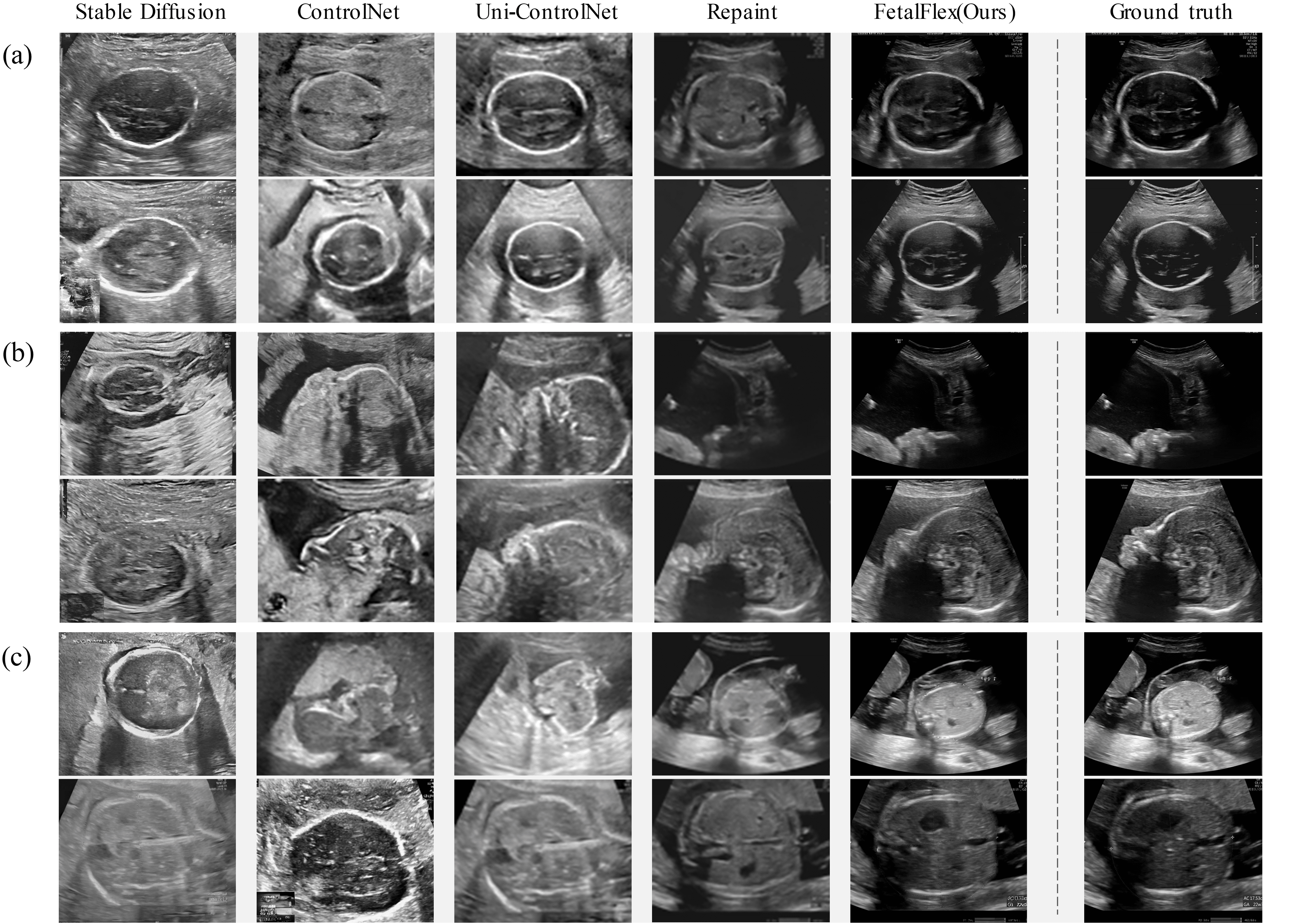}
    \caption{Examples of three classes of synthetic fetal US images generated using five comparison methods, while the sixth column shows Ground truth.}
    \label{c11}
\end{figure*}

\begin{table*}[t]
\centering
\caption{Quantitative results for 512x512 fetal US image synthesis on three fetal US planes. nR, naive Repaint; MA, full modal architecture design. The default sample method is DDIM, which step is set to 50. Statistical significance was tested with $p << .001$.}
\renewcommand{\arraystretch}{1.05} %
\resizebox{\textwidth}{!}{
\begin{tabular}{l|cccc|cccc|cccc}
\toprule
\multirow{2}{*}{Setting} & \multicolumn{4}{c|}{Thalamic transverse plane}                      & \multicolumn{4}{c|}{Facial sagittal plane}                          & \multicolumn{4}{c}{Upper abdominal transverse plane}                \\
                         & PSNR↑           & MS-SSIM↑       & FID↓            & LPIPS↓         & PSNR↑           & MS-SSIM↑       & FID↓            & LPIPS↓         & PSNR↑           & MS-SSIM↑       & FID↓            & LPIPS↓         \\ \midrule 
Baseline                 & 11.066          & 0.332          & 145.919         & 0.554          & 8.853           & 0.254          & 159.096         & 0.567          & 9.271           & 0.260          & 157.462         & 0.583          \\
Baseline+nR              & 19.294          & 0.762          & 73.629          & 0.198          & 24.817          & 0.912          & 61.048          & 0.119          & 19.027          & 0.786          & 76.992          & 0.220          \\
Baseline+nR+MA           & 20.087          & 0.765          & 58.479          & 0.184          & 25.025          & \textbf{0.914} & 52.723          & 0.112          & 20.783          & 0.796          & 57.561          & 0.164          \\
FetalFlex (Ours)        & \textbf{20.538} & \textbf{0.766} & \textbf{54.485} & \textbf{0.178} & \textbf{25.263} & \textbf{0.914} & \textbf{50.734} & \textbf{0.105} & \textbf{21.284} & \textbf{0.803} & \textbf{52.157} & \textbf{0.159} \\ \bottomrule
\end{tabular}
}
\label{a4}
\end{table*}

\begin{table*}[t]
\centering
\caption{Quantitative results for 512x512 fetal US image synthesis using different sampling strategy. The best results are in bold and the second best results are underlined. Statistical significance was tested with $p << .001$.}
\renewcommand{\arraystretch}{1.04} %
\resizebox{\textwidth}{!}{
\begin{tabular}{l|cccc|cccc|cccc}
\toprule
\multirow{2}{*}{Setting} & \multicolumn{4}{c|}{Thalamic transverse plane}                          & \multicolumn{4}{c|}{Facial sagittal plane}                              & \multicolumn{4}{c}{Upper abdominal transverse plane}                
\\
& PSNR↑           & MS-SSIM↑       & FID↓            & LPIPS↓   
& PSNR↑           & MS-SSIM↑       & FID↓            & LPIPS↓   
& PSNR↑           & MS-SSIM↑       & FID↓            & LPIPS↓  
\\ \midrule
Task-specific training        
& 19.986          & 0.749           & 65.051        & 0.186           
& 24.934          & 0.912           & 50.943        & \underline {0.103}           
& 12.883          & 0.285          & 62.909         & 0.512     
\\
DDIM (50 steps)           
& 20.538          & 0.766          & 54.485          & 0.178          
& 25.263          & 0.914          & 50.734          & 0.105          
& 21.284          & 0.803          & 52.157          & 0.159          
\\
DDIM (100 steps)          
& 20.692          & 0.798          & 53.883          & 0.164          
& 25.197          & 0.920          & 50.256          & 0.106          
& 21.194          & 0.864          & 51.981          & \underline {0.157}          
\\
SSA (Ours)                      
& \textbf{22.010} & \underline {0.823}          & \textbf{51.631} & \underline {0.174}          
& \textbf{26.274} & \textbf{0.937} & \underline {50.213}          & \textbf{0.102} 
& \textbf{22.774} & \textbf{0.869} & \underline {51.465}          & \underline {0.157}    
\\
SSA+OSM (Ours)                  
& \underline {21.817}          & \textbf{0.835} & \underline {51.951}          & \textbf{0.173} 
& \underline {25.858}          & \underline {0.936}          & \textbf{49.595} & \underline {0.103}          
& \underline {21.979}          & \underline {0.869}          & \textbf{50.935} & \textbf{0.156}     
\\ \bottomrule
\end{tabular}
}
\label{a2}
\end{table*}

\begin{figure*}[!h]
    \centering
    \includegraphics[width=0.99\linewidth]{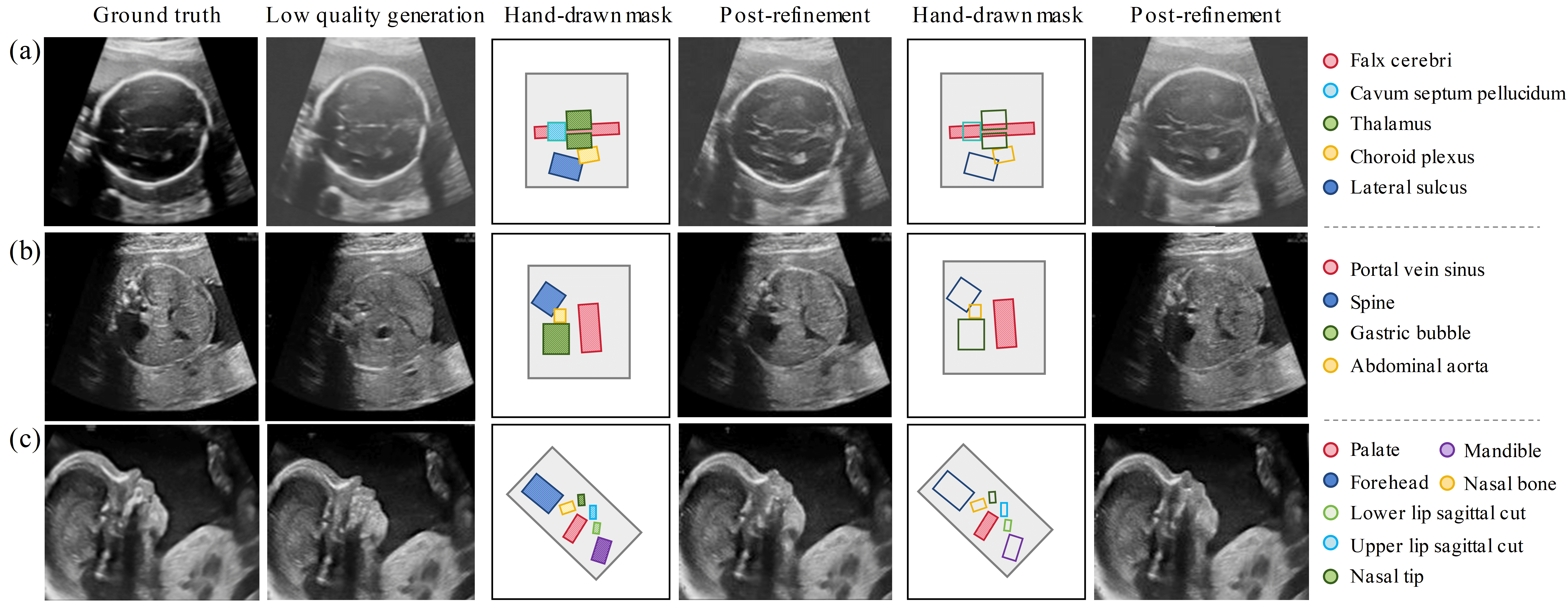}\\ 
    \caption{Visualization of the OSM sampling for post-processing refinement.}
    \label{post}
\end{figure*}

\subsection{Implementation Details}
We trained the FetalFlex to generate controllable fetal US images conditioned on layout maps, ROI masks, and global text prompts. 
FetalFlex utilizes a conditional LDM 1.5~\citep{high} as the core framework, which was trained on four NVIDIA 3090 GPUs. We optimized the model with a total of 1000 timesteps (T) and a linear noise schedule where the noise coefficient $\beta_{t}$ ranged from $\beta_{1}$ = $10^{-4}$ to $\beta_{T}$ = 0.02. 
The training process lasted for 120 epochs and used a learning rate of 1e-5 to generate images with a resolution of 512x512 pixels.

\subsection{Comparison Models and Evaluation Metrics}
To the best of our knowledge, there is no existing research that performed the generation of fetal US planes. 
Therefore, to facilitate comparison, we implemented baseline synthesis methods developed for other application domains and adapted them for generating multiple planes of fetal US images.\\
\textbf{Stable Diffusion} \citep{diffusion}.This method generates images through a series of denoising steps. However, SD struggles to adapt to the domain of medical image synthesis owing to the lack of condition control.\\
\textbf{ControlNet} \citep{controlnet}. This introduces an additional network branch that incorporates external control signals, which enhances the process by enabling the generated images to more accurately align with the input conditions.\\
\textbf{UnicontrolNet} \citep{uni}. Builds upon the ControlNet and is capable of handling a diverse range of mixed control conditions, including local controls and global controls.\\
\textbf{Repaint} \citep{3-01}. This approach adds a mask to the original image and guides the sampling process through the unmasked regions to ensure image consistency. However, it generates results that tend to be learned data distributions that are not user controllable.

\textbf{Evaluation Metrics}. We conducted a comprehensive evaluation of the quality of the synthesized images, encompassing both quantitative and qualitative aspects. Regarding quantitative assessment, we selected several commonly used image quality evaluation metrics, including MS-SSIM \citep{ssim}, PSNR \citep{PSNR}, FID \citep{FID}, and LPIPS \citep{LPIPS}. MS-SSIM measures structural similarity across multiple scales, whereas PSNR is an important metric for evaluating peak signal-to-noise ratio in images. FID calculates the distribution distance between generated images and real images in the deep feature space, and LPIPS focuses on the visual similarity as perceived by humans. Furthermore, we included traditional metrics such as Accuracy, Recall, Precision, F1-score, and AUC to validate the effectiveness of the synthesized images in two downstream applications.

\section{Results and Analysis}
We conducted a series of experiments to investigate and assess the proposed FetalFlex model, including comparative experiments, ablation studies, quantitative evaluations, downstream tasks, and reader studies.
Lastly, we provided visualization results to better discuss the performance and controllability of the generated model.

\subsection{Comparison Study}
In this section, we compared the performance of the proposed FetalFlex model against four baseline algorithms on an internal multi-center test set. Additionally, we explored the impact of incorporating control conditions within the diffusion model framework on the synthesis quality. Table~\ref{c1} provides the quantitative evaluation results, and Fig.~\ref{c11} presents visual examples.

We adapted the input control conditions to suit each model's architecture owing to structural differences among the models. Quantitatively, FetalFlex consistently outperforms all baseline methods, achieving the highest PSNR (23.686) and MS-SSIM (0.876), and the lowest FID (51.103) and LPIPS (0.144) scores. These metrics indicated that FetalFlex produces images with higher fidelity, greater structural similarity to the GT, and better perceptual quality than other methods.
The qualitative results in Fig. \ref{c11} further corroborated these findings. Images generated by FetalFlex exhibited finer details, better contrast, and fewer artifacts, which are crucial for accurate anatomical analysis. In contrast, SD generates noisy images with US texture but lacks semantic information. ControlNet and Uni-ControlNet, which incorporate layout maps and a combination of layout maps with Canny edges, respectively, struggled to adapt to the multi-class data distribution in small-scale datasets with confusing text embedding. The Repaint method was based on FetalFlex's weight and architecture; however, its performance remained inferior to that of FetalFlex, particularly in maintaining anatomical consistency, due to the absence of control conditions. In summary, the proposed FetalFlex model achieved superior synthesis results in both metrics and visual quality.

\subsection{Ablation Study}
In this study, FetalFlex was improved regarding two key aspects: model architecture components and sampling strategy.
We implemented two parts of ablation experiments in Sections \ref{a1-11} and \ref{a1-12} to quantify the specific contributions of these improvements to overall performance.

\subsubsection{Ablation on Model Architecture}
\label{a1-11}
We conducted an ablation study on a 512x512 fetal US image synthesis task across three US planes to comprehensively evaluate the contribution of several components in FetalFlex.
We compared the results of the following three methods: Baseline, Baseline+nR, Baseline+nR+MA, and proposed FetalFlex. The Baseline method uses only the backbone structure of ControlNet, while Baseline+nR incorporates the inpainting strategy (Section \ref{3.3.2}), and Baseline+nR+MA employs the full model architecture (Section \ref{3.3.1} ). Finally, our proposed FetalFlex represents the last model architecture with ensemble loss (Section \ref{3.3.4}). We compared the performance by incrementally integrating key components into the baseline model.

The quantitative results in Table \ref{a4} demonstrate that adding the inpainting strategy to the baseline significantly improves all metrics. This is because effectively learning complex image features and details on a small number of training samples is difficult for the model. The model can utilize the conditional information to perform more accurate image restoration with this strategy, thus enhancing its feature learning ability and generation quality. Further enhancements are observed with the pre-alignment module, pre-trained Bert, with a notable improvement in the FID score. The last row presents the full model architecture with ensemble loss. These findings underscore the effectiveness of these components in enhancing the US image synthesis process.

\subsubsection{Ablation on Sampling Strategy}
\label{a1-12}
In this section, we present experimental results on different sampling strategies. To evaluate whether the text-prompted universal framework effectively enables the model to learn common features of US images, we first compared the outcomes of training independently on three fetal planes versus training using the FetalFlex. 
As shown in Table \ref{a2}, the results indicate that the model fails to achieve a comprehensive understanding of the overall structure and features of US images with separate training, resulting in relatively lower performance, particularly for the upper abdominal plane.
In contrast, joint training across multiple planes using FetalFlex significantly improved the overall quality and consistency of the generated images.

Furthermore, to validate the effectiveness of our proposed SSA and OSM sampling methods, we compared the results of the following approaches: DDIM (steps = 50), DDIM (steps = 100), SSA, and SSA+OSM. 
As shown in rows 2 to 5 of Table \ref{a2}, simply increasing the number of sampling steps did not lead to stable improvements in the generated results. 
In contrast, introducing SSA sampling into the inference process significantly enhanced image quality and detail preservation, demonstrating its superiority in image generation. Moreover, the SSA+OSM combination, which primarily addresses abnormal synthesis at the anatomical structure level, showed substantial improvement in generating consistent results. When evaluating the two-stage sampling strategy, we subjectively selected the organ with the largest area as the organ-level mask. Additionally, the evaluation results achieved metrics similar to SSA sampling since OSM does not involve user intervention.

Notably, the OSM sampling strategy can be employed as a post-processing technique to generate high-quality images from computationally less expensive (lower-quality) ones, thereby refining the flawed outputs produced by other networks.
As illustrated in Fig. \ref{post}, in the first row, user-drawn masks were applied to refine one or more organ regions in the original images generated by Uni-ControlNet, guided by text prompts. The second row demonstrates improvements in anatomical structure and texture in low-quality US images reproduced by FetalFlex. In the last row, for synthesizing positive samples, precise control over local anatomical structures was achieved by modifying only the regions with structural deficiencies while leaving other areas unchanged. In conclusion, our method effectively refined all masked regions in a single sampling process. 
This strategy ensures targeted improvements without compromising the overall integrity of the image.

\begin{figure}[t]
    \centering
    \includegraphics[width=0.99\linewidth]{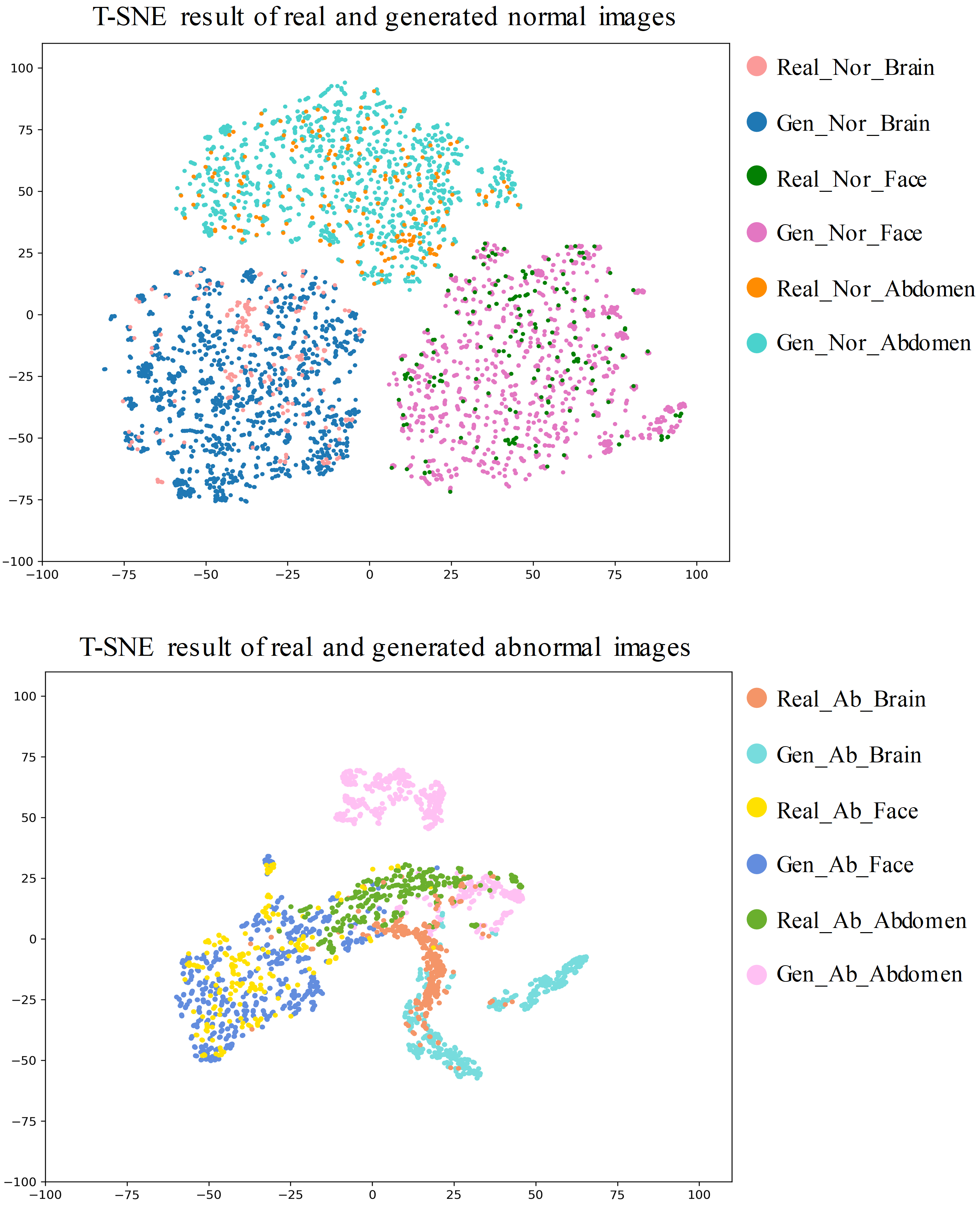}
    \caption{An overview of the feature embeddings of real/generated normal fetal US images and real/generated abnormal fetal US images through t-SNE visualization.}
    \label{plot}
\end{figure}

\subsection{Quantitative Evaluation}
The t-SNE plots in Fig. \ref{plot} illustrate a comparison of feature distributions between real and generated fetal US images, both normal and abnormal. The feature embeddings are extracted using a pre-trained ResNet50 model. Each color represents different classes of fetal US images. The first t-SNE plot compares the feature distributions of real and generated normal images. The results demonstrated that the features of the generated images exhibit a clear clustering pattern, closely aligning with the distribution of the real normal US images on the same scale. This finding confirmed that the FetalFlex model effectively augments the dataset by generating image-label pairs that maintain category consistency while preserving feature diversity. In contrast, the second plot presents the feature distributions of real and generated abnormal images. The inter-class distribution of the abnormal images showed some variability; nevertheless, the overall distribution remained similar to that of the normal images. This is likely due to the consistent anatomical structure of normal fetuses, while that of abnormal fetuses tends to be more diverse and complex. Abnormal US images are not entirely distinct from normal anatomy; therefore, a degree of structural overlap exists between them. These results indicate that the FetalFlex model can simulate abnormalities while retaining certain aspects of the normal anatomical background.

\subsection{Downstream Experiments}
To provide a more comprehensive evaluation of the normal and abnormal samples synthesized by FetalFlex, we conducted two downstream experiments: classification and anomaly detection. This extended assessment aimed to further validate the practicality and effectiveness of FetalFlex in real-world medical imaging application scenarios.

\begin{table*}[!h]
    \centering
    \caption{Accuracy metric comparison of different networks trained on real, mixed real-generated, and generated data, tested exclusively on real data. The best results are in bold, and the second-best are underlined. Aug represents the traditional data augmentation methods. All models are trained from sketch.}
    \vspace{1mm}
\resizebox{1\textwidth}{!}{
\begin{tabular}{l|cccccc|c}
\toprule
\multirow{2}{*}{Training Dataset} & VGG16                                            & ResNet50                                         & DenseNet121                                      & Vit-B                                            & ResNeXt                                          & RegNet                                            & \multirow{2}{*}{Average} \\
                                  & \multicolumn{1}{l}{\citet{vgg}} & \multicolumn{1}{l}{\citet{resnet}} & \multicolumn{1}{l}{\citet{dense}} & \multicolumn{1}{l}{\citet{vit}} & \multicolumn{1}{l}{\citet{resnext}} & \multicolumn{1}{l|}{\citet{reg}} &                          \\ \midrule
100\%Real            & 0.91                      & 0.87          & 0.90          & 0.73          & 0.83          & 0.70          & 0.823                       \\
100\%Gen             & 0.85                      & 0.85          & 0.89          & 0.72          & 0.79          & 0.71          & 0.802                       \\
50\%Real+50\%Gen     & \underline {0.92}                & 0.92          & \underline {0.96}    & 0.74          & \underline {0.88}    & 0.70          & 0.853                       \\
100\%Real+Aug        & 0.91                      & \underline {0.95}    & 0.90          & \underline {0.77}    & 0.83          & \underline {0.84}    & 0.867                       \\
100\%Gen+Aug         & 0.85                      & 0.93          & \underline {0.96}    & \textbf{0.78} & \textbf{0.94} & \textbf{0.87} & \underline {0.888}                 \\
50\%Real+50\%Gen+Aug & \textbf{0.96}             & \textbf{0.96} & \textbf{0.98} & \underline {0.77}    & 0.84          & \textbf{0.87} & \textbf{0.897}              \\ \bottomrule
\end{tabular}}
\label{t33344}
\end{table*}

\begin{table*}[!h]
    \centering
    \caption{Precision metric comparison of different networks trained on real, mixed real-generated, and generated data, tested exclusively on real data. The best results are in bold, and the second-best are underlined. Aug represents the traditional data augmentation methods. All models are trained from sketch.}
    \vspace{1mm}
\resizebox{1\textwidth}{!}{
\begin{tabular}{l|cccccc|c}
\toprule
\multirow{2}{*}{Training Dataset} & VGG16                                            & ResNet50                                         & DenseNet121                                      & Vit-B                                            & ResNeXt                                          & RegNet                                            & \multirow{2}{*}{Average} \\
                                  & \multicolumn{1}{l}{\citet{vgg}} & \multicolumn{1}{l}{\citet{resnet}} & \multicolumn{1}{l}{\citet{dense}} & \multicolumn{1}{l}{\citet{vit}} & \multicolumn{1}{l}{\citet{resnext}} & \multicolumn{1}{l|}{\citet{reg}} &                          \\ \midrule
100\%Real            & 0.912          & 0.873          & 0.901          & 0.739          & 0.831          & 0.721          & 0.830          \\
100\%Gen             & 0.851          & 0.853          & 0.892          & 0.721          & 0.801          & 0.733          & 0.809          \\
50\%Real+50\%Gen     & \underline {0.920}    & 0.920          & 0.961          & 0.748          & 0.880          & 0.706          & 0.856          \\
100\%Real+Aug        & 0.910          & \underline {0.951}    & 0.903          & 0.782          & \underline {0.930}    & 0.842          & \underline {0.886}    \\
100\%Gen+Aug         & 0.852          & 0.931          & \underline {0.962}    & \textbf{0.790} & 0.841          & \underline {0.876}    & 0.875          \\
50\%Real+50\%Gen+Aug & \textbf{0.960} & \textbf{0.961} & \textbf{0.980} & \underline {0.787}    & \textbf{0.942} & \textbf{0.882} & \textbf{0.919} \\ \bottomrule
\end{tabular}}
\label{precision}
\end{table*}

\begin{table*}[!h]
    \centering
    \caption{Recall metric comparison of different networks trained on real, mixed real-generated, and generated data, tested exclusively on real data. The best results are in bold, and the second-best are underlined. Aug represents the traditional data augmentation methods. All models are trained from sketch.}
    \vspace{1mm}
\resizebox{1\textwidth}{!}{
\begin{tabular}{l|cccccc|c}
\toprule
\multirow{2}{*}{Training Dataset} & VGG16                                            & ResNet50                                         & DenseNet121                                      & Vit-B                                            & ResNeXt                                          & RegNet                                            & \multirow{2}{*}{Average} \\
                                  & \multicolumn{1}{l}{\citet{vgg}} & \multicolumn{1}{l}{\citet{resnet}} & \multicolumn{1}{l}{\citet{dense}} & \multicolumn{1}{l}{\citet{vit}} & \multicolumn{1}{l}{\citet{resnext}} & \multicolumn{1}{l|}{\citet{reg}} &                          \\ \midrule
100\%Real            & 0.909          & 0.870          & 0.900          & 0.730          & 0.830          & 0.701          & 0.823          \\
100\%Gen             & 0.851          & 0.851          & 0.890          & 0.720          & 0.792          & 0.712          & 0.803          \\
50\%Real+50\%Gen     & \underline {0.920}    & 0.920          & 0.960          & 0.738          & 0.881          & 0.701          & 0.853          \\
100\%Real+Aug        & 0.910          & \underline {0.950}    & 0.901          & 0.769          & \underline {0.931}    & 0.841          & \underline {0.884}    \\
100\%Gen+Aug         & 0.852          & 0.931          & \underline {0.961}    & \textbf{0.781} & 0.842          & \underline {0.870}    & 0.873          \\
50\%Real+50\%Gen+Aug & \textbf{0.959} & \textbf{0.960} & \textbf{0.980} & \underline {0.769}    & \textbf{0.941} & \textbf{0.871} & \textbf{0.913} \\ \bottomrule
\end{tabular}}
\label{recall}
\end{table*}

\begin{table*}[!h]
    \centering
    \caption{F1-score metric comparison of different networks trained on real, mixed real-generated, and generated data, tested exclusively on real data. The best results are in bold, and the second-best are underlined. Aug represents the traditional data augmentation methods. All models are trained from sketch.}
    \vspace{1mm}
\resizebox{1\textwidth}{!}{
\begin{tabular}{l|cccccc|c}
\toprule
\multirow{2}{*}{Training Dataset} & VGG16                                            & ResNet50                                         & DenseNet121                                      & Vit-B                                            & ResNeXt                                          & RegNet                                            & \multirow{2}{*}{Average} \\
                                  & \multicolumn{1}{l}{\citet{vgg}} & \multicolumn{1}{l}{\citet{resnet}} & \multicolumn{1}{l}{\citet{dense}} & \multicolumn{1}{l}{\citet{vit}} & \multicolumn{1}{l}{\citet{resnext}} & \multicolumn{1}{l|}{\citet{reg}} &                          \\ \midrule
100\%Real            & 0.910          & 0.868          & 0.899          & 0.732          & 0.830          & 0.702          & 0.824          \\
100\%Gen             & 0.848          & 0.849          & 0.889          & 0.712          & 0.784          & 0.703          & 0.798          \\
50\%Real+50\%Gen     & \underline {0.920}    & 0.919          & \underline {0.960}    & 0.733          & 0.880          & 0.700          & 0.852          \\
100\%Real+Aug        & 0.911          & \underline {0.949}    & 0.900          & 0.760          & \underline {0.929}    & 0.840          & \underline {0.882}    \\
100\%Gen+Aug         & 0.848          & 0.930          & 0.959          & \textbf{0.780} & 0.835          & \textbf{0.872} & 0.871          \\
50\%Real+50\%Gen+Aug & \textbf{0.960} & \textbf{0.960} & \textbf{0.980} & \underline {0.771}    & \textbf{0.940} & \underline {0.869}    & \textbf{0.913} \\ \bottomrule
\end{tabular}}
\label{f1}
\end{table*}

\subsubsection{Fetal US Planes Classification Task}
We found that most studies only expanded the original dataset with the synthetic results of the generated model and evaluated the enhancement effect of the expanded dataset of the training set for downstream experiments. While this approach often leads to positive results owing to the increased data quantity, we argue that validation methods based on data expansion do not fully assess the validity of synthetic results, ignoring the focus of the generative field's attention, which is whether synthetic samples are promising substitutes for real data.

\begin{figure*}[!h]
    \centering
    \includegraphics[width=1\linewidth]{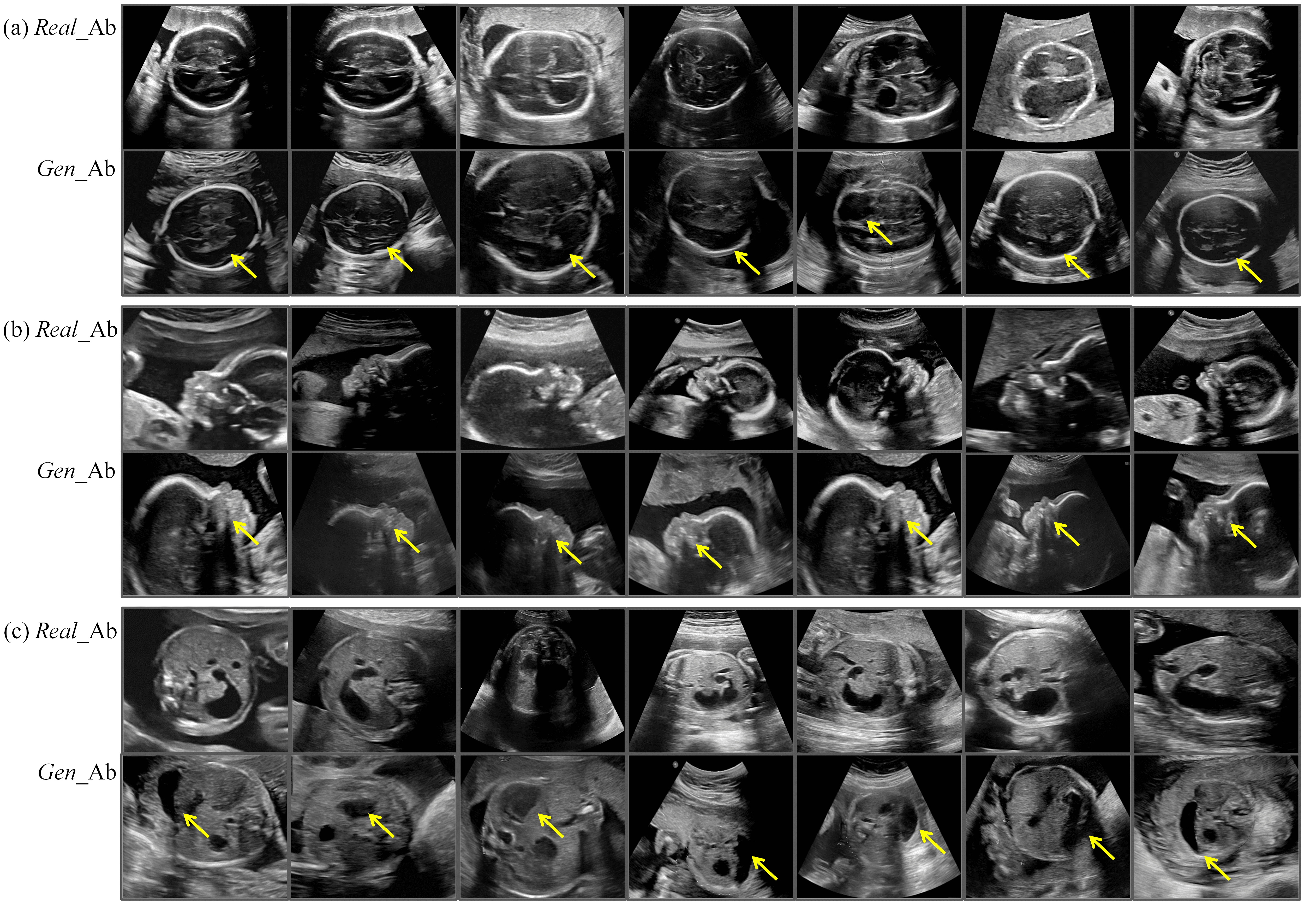}
    \caption{Real and generated abnormal fetal US images. (a) (b) (c) represent three different fetal US datasets. The first row displays a subset of real cases of hydrocephalus, the third row shows a subset of real cases of cleft lip and palate, and the fifth row presents a subset of real cases of duodenal atresia or stenosis. The second, fourth, and sixth rows correspond to abnormal samples generated by FetalFlex, where the yellow arrows indicate the anatomical structures that generate results similar to the real anomaly samples. Real\_Ab: Real abnormal US images; Gen\_Ab: Generated abnormal US images.}
    \label{vis999}
\end{figure*}

We constructed six training sets with a fixed sample number but different data proportions to verify the authenticity of the synthetic data generated by FetalFlex.
Specifically, we examined the impact of using 600 real samples (100\% Real), 300 real samples combined with 300 synthetic samples generated by FetalFlex (50\% Real + 50\% Gen), and 600 fully synthetic samples generated by FetalFlex (100\% Gen) on the ability of deep learning models. Additionally, we also assessed the impact of traditional data augmentation methods, like horizontal flip, hue = 0.5, light = 0.5. These training datasets were used to train six classical deep learning networks to classify the category of fetal US images: thalamic transverse plane, facial sagittal plane, and upper abdominal transverse plane. The test set only consists of 100 balanced real US images from clinical scenarios. With this experimental design, we provided a comprehensive analysis of the substantial contribution of synthetic samples generated by FetalFlex in the classification task. 

Tables \ref{t33344}, \ref{precision}, \ref{recall} and \ref{f1} present the accuracy, precision, recall, and F1-score of six deep learning models on the test set under different training set configurations. The results indicate that, while keeping the training set size constant, models trained solely on synthetic data performed slightly worse than those trained on real data. However, simple traditional data augmentation methods can narrow this performance gap. Notably, regarding average accuracy, models trained exclusively on synthetic data even outperformed those trained on real data, showing an improvement of 2.1\%.
Further analysis revealed that using a mixed training set with 50\% real data and 50\% generated data significantly enhanced the overall model performance, achieving the best results across all four classification metrics. We hypothesized that the samples generated by FetalFlex enrich the training set by providing greater diversity, allowing the model to learn more generalized class features, which better equips it to handle the heterogeneity of test sets composed of real data. This demonstrates that the data distribution of FetalFlex-generated data aligns closely with that of real data, ensuring the effectiveness and reliability of synthetic data in downstream tasks. Moreover, FetalFlex offers a viable solution to the challenges of data privacy and accessibility.

\subsubsection{Anomaly Detection Task}
\label{c-12}
We conducted anomaly detection experiments to evaluate the reliability of normal and abnormal samples synthesized by FetalFlex on three planes. We shifted and altered the layout maps and corresponding global prompts to generate fetal US samples with abnormal features, ensuring a rigorous assessment of the consistency and validity of the image-label pairs and achieving the transformation of ID normal data to OOD abnormal data. 

It is important to note that in actual clinical prenatal screening, the types of fetal anomalies are highly diverse, involving multiple anatomical structures and physiological systems. This complexity causes challenges in retrospective data collection owing to the difficulty in covering all possible types of anomalies. Therefore, the retrospectively collected multi-center real abnormal dataset has certain limitations regarding the representation of anomalies across different fetal planes. In detail, the dataset for the transverse thalamic plane contains only cases of hydrocephalus, the facial sagittal plane dataset includes only fetuses with cleft lip and palate, and the upper abdominal transverse plane dataset is limited to cases of duodenal atresia or stenosis. For this reason, we instructed FetalFlex to perform a high-quality synthesis of abnormalities for the anatomical and structural manifestation characteristics of the abovementioned diseases. We visualized the real clinical anomaly data collected alongside the abnormal samples generated by FetalFlex, as shown in Fig. \ref{vis999}. By adjusting the control conditions, FetalFlex was able to generate US images that closely resemble the abnormal in real cases, allowing our proposed method to achieve cross-domain generation at the image-level.

\begin{table*}[t]
    \centering
    \caption{OOD task metrics comparison for pretrained ResNet50 \citet{resnet} finetuned on Real unbalanced data, mixed Real-Generated Normal data, and mixed Real-Generated Normal and Abnormal data, tested exclusively on class-balanced real dataset. The best results are in bold and the second-best are underlined. Real: Real unbalanced normal and abnormal data; GN: Generated Normal data; GAb: Generated Abnormal data.}
\vspace{1mm}
\renewcommand{\arraystretch}{1.25} 
\resizebox{\textwidth}{!}{
\begin{tabular}{l|ccccc|ccccc|ccccc}
\toprule
\multirow{2}{*}{Training Dataset} & \multicolumn{5}{c|}{Thalamic transverse plane}                                          & \multicolumn{5}{c|}{Facial sagittal plane}                                              & \multicolumn{5}{c}{Upper abdominal transverse plane}                                    \\
                                  & Precision       & Accuracy        & Recall          & F1-score        & AUC             & Precision       & Accuracy        & Recall          & F1-score        & AUC             & Precision       & Accuracy        & Recall          & F1-score        & AUC             \\ \midrule
Real                          & 0.919          & 0.802          & 0.667          & 0.773          & 0.952          & \underline {0.897}          & 0.737          & 0.531          & 0.667          & \underline {0.894}          & 0.703          & 0.768          & \underline {0.918}          & 0.797          & 0.787          \\
Real+GN                 & \underline {0.958}          & \underline {0.931}          & \underline {0.902}          & \underline {0.929}          & \textbf{0.986} & 0.822          & \underline {0.798}          & \textbf{0.755} & \underline {0.787}          & \textbf{0.906} & \underline {0.927}          & \underline {0.859}          & 0.776          & \underline {0.844}          & \underline {0.952}          \\
Real+GN+GAb         & \textbf{0.959} & \textbf{0.941} & \textbf{0.922} & \textbf{0.940} & \underline {0.970}          & \textbf{0.947} & \textbf{0.848} & \underline {0.735}          & \textbf{0.828} & 0.884          & \textbf{0.978} & \textbf{0.949} & \textbf{0.918} & \textbf{0.947} & \textbf{0.985} \\ \bottomrule
\end{tabular}}
\label{t44}
\end{table*}

\begin{table}[!h]
\centering
\caption{Dataset partitioning for anomaly detection task.}
\renewcommand{\arraystretch}{1} %
\resizebox{0.49\textwidth}{!}{
\begin{tabular}{l|cccc}
\toprule
OOD Dataset Setting & Real\_Nor & Gen\_Nor & Real\_Ab & Gen\_Ab \\ \midrule
Training set \#1 (Real)        & 400       & -        & 20          & -          \\
Training set \#2 (Real+GN)        & 400       & 400      & 20          & -          \\
Training set \#3 (Real+GN+GAb)        & 400       & 400      & 20          & 400        \\
\midrule
Validation set           & 50        & -        & 50          & -          \\
Test set                 & 51        & -        & 50          & -          \\ \bottomrule
\end{tabular}}
\label{a21}
\end{table}

In the experimental setup, we compared the performance of the pre-trained ResNet50 model across three different training datasets on three types of fetal US images. 
As shown in Table \ref{a21}, firstly, we constructed a dataset with 400 normal and 20 abnormal samples to simulate the scarcity of abnormal samples in real clinical scenarios. This baseline training set was long-tailed, with a class imbalance ratio of 20:1, presenting a significant challenge. Next, we synthesized 400 normal samples using FetalFlex, which were added to the baseline dataset to form the second training set (Real+GN). The goal was to enable the model to learn the anatomical structural features of normal samples, thereby distinguishing abnormal samples in the test set. Finally, we guided FetalFlex to synthesize large volumes of abnormal data, creating a mixed dataset (Real+GN+GAb) to help ResNet50 learn the semantic-level feature distribution differences between normal and abnormal samples. The trained ResNet50 was tested on a balanced real dataset covering 51 real normal samples and 50 real abnormal samples. We evaluated its performance using five metrics: precision, accuracy, recall, F1-score, and AUC across three types of fetal US.

As shown in Table \ref{t44}, we gradually added generated normal (GN) and generated abnormal (GAb) data during training to evaluate the model boosting effect. As shown in the first row, when training with only unbalanced real data, the downstream model's performance was moderate, particularly in the facial sagittal plane, where the F1-score and recall were 0.667 and 0.531, respectively. Therefore, we added 400 normal samples generated by FetalFlex, and the false positive rate was greatly improved, particularly on the thalamic transverse plane and facial sagittal plane. We believe that generating more diverse normal samples allowed the anomaly detection model to learn richer features of normal cases, thereby enhancing its ability to distinguish between normal and abnormal samples.

\begin{figure*}[!ht]
    \centering
    \includegraphics[width=1\linewidth]{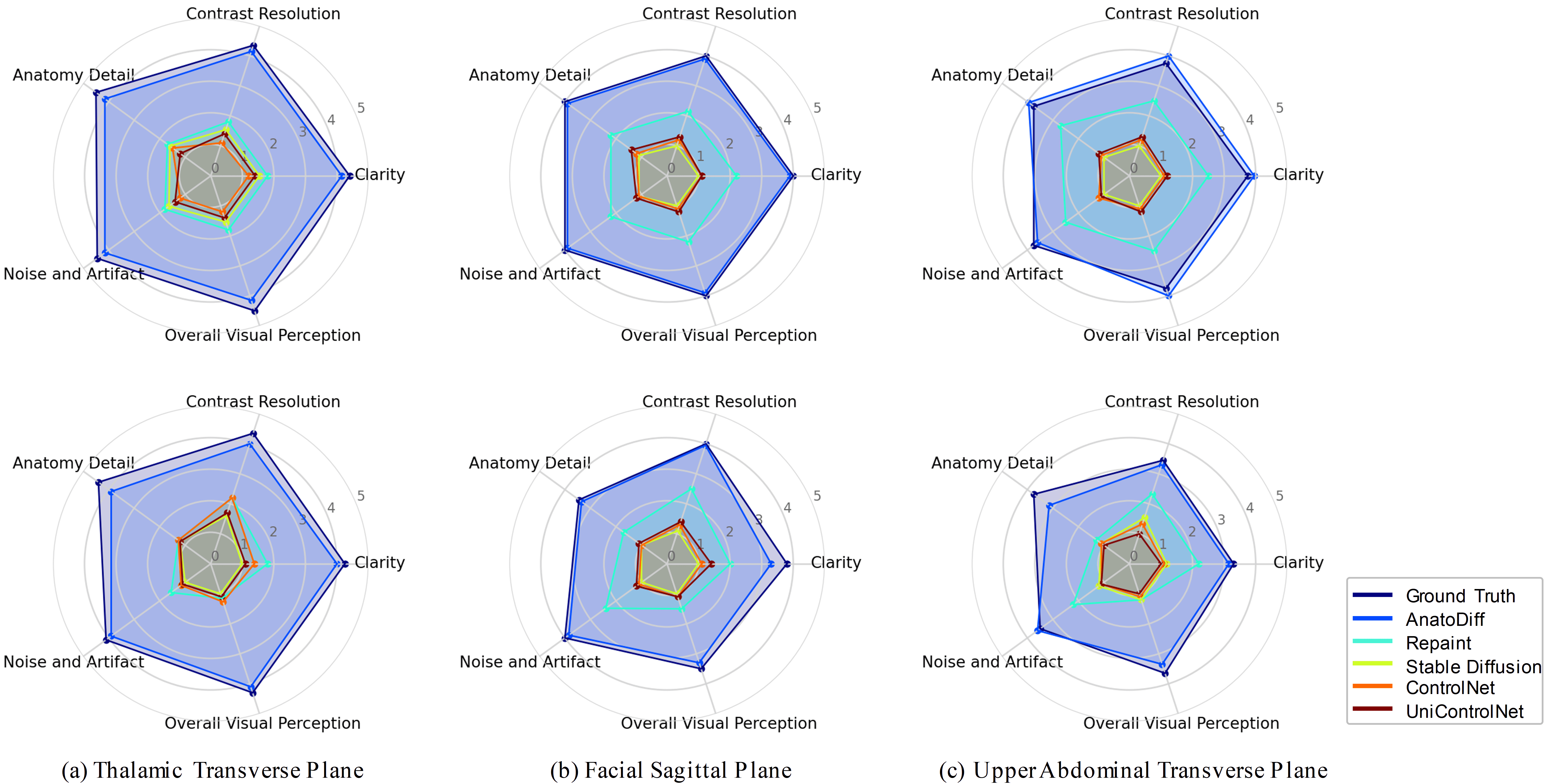}\\ 
    \caption{Comparison of generated image quality ratings (on a scale of 1 - 5) by two senior asset fetal US radiologists across five different methods and the GT images. The first row represents the average score of the first radiologist, the second row represents the average score of the second radiologist, and the first, second, and third columns represent the three image categories: Thalamic US images, Facial US images, and Abdominal US images.}
    \label{reader}
\end{figure*}

The performance of all metrics consistently improved when the training dataset included both GN and GAb data, resulting in the highest F1-score, precision, and accuracy across all three fetal US planes. The model achieved a substantial increase in precision and accuracy in the thalamic transverse plane, with a significant reduction in the false positive rate. Similarly, the model achieved a high F1-score of 0.947 in the upper abdominal transverse plane, reflecting a balanced trade-off between recall (0.918) and precision (0.978), effectively minimizing false positives while maintaining a high true positive rate. We believe that generating more diverse normal samples allowed the anomaly detection model to learn richer features of normalities, thereby enhancing its ability to distinguish between normal and abnormal samples. The most notable improvements occur in the challenging facial sagittal plane, where subtle structural changes in cases of cleft palate caused difficulty for the model to learn fine distinctions. The introduction of synthesized data reduced false positives and improved the detection rate. These results indicated that the abnormal samples synthesized by FetalFlex contain strong image characteristics of abnormalities, guiding the downstream model to learn a wider range of abnormal fetal US patterns. FetalFlex helps balance highly imbalanced datasets by increasing the number of abnormal samples, thereby further enhancing model performance.

\begin{figure*}[!h]
    \centering
    \includegraphics[width=1\linewidth]{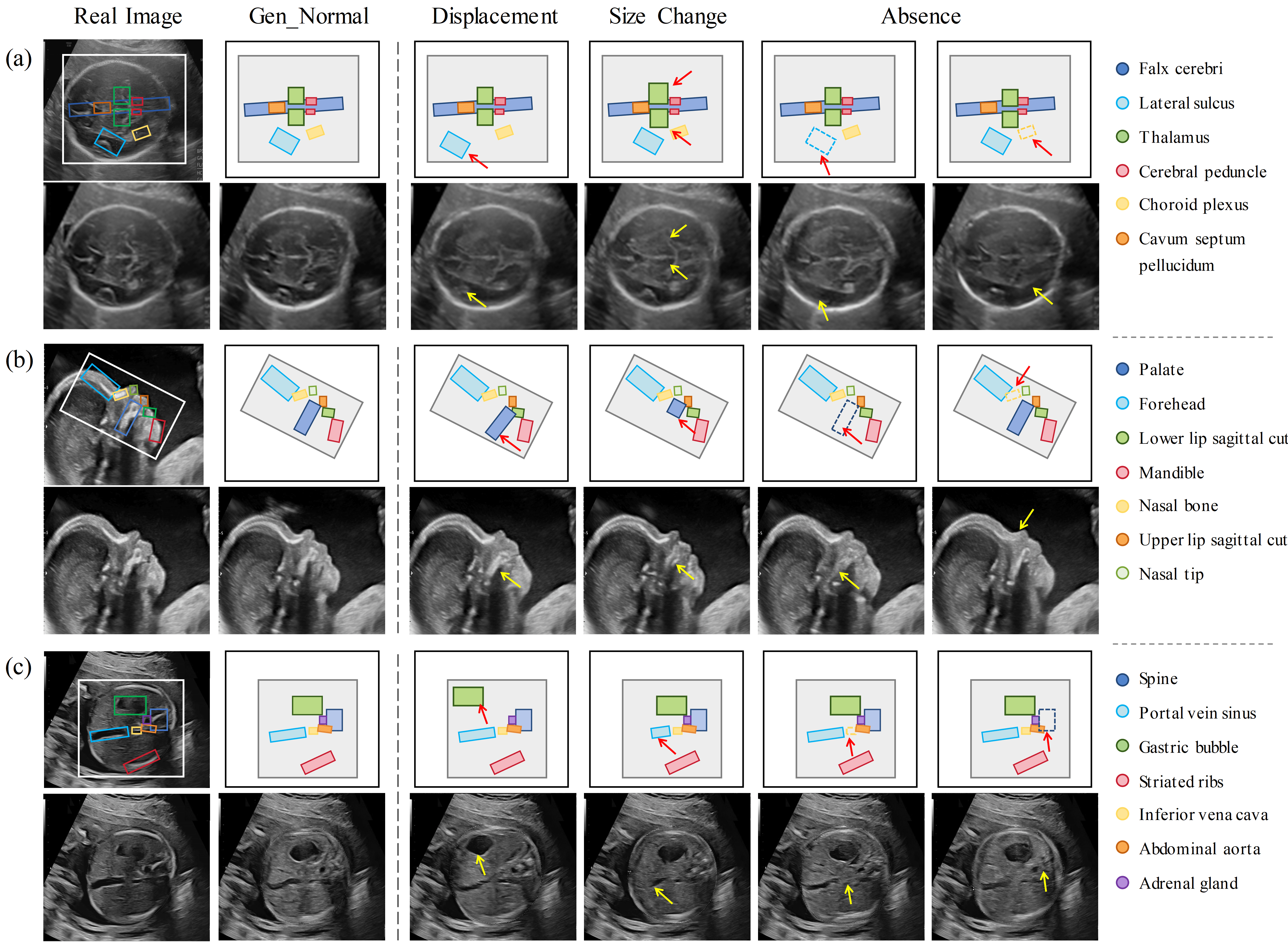}
    \caption{Samples of normal and abnormal images generated by the proposed FetalFlex under anatomical structure-related layout maps, where the red arrows point out the anatomical structure change in the user-edited layout maps, and the yellow arrows highlight the anatomical structure change in the corresponding synthesized images.}
    \label{vis}
\end{figure*}

\subsection{Reader Study}
To thoroughly evaluate the synthesized results from a clinical perspective, we randomly selected 360 images from three planes of fetal US synthetic results by five comparison methods, also including corresponding ground truth (GT) images. 
We obtained 20 images for each US plane category, resulting in 60 images per generation method. Two senior radiologists were asked to review these anonymized images and rate them based on five assessment criteria: clarity, brightness and contrast, anatomical detail, noise level and artifacts, and subjective image quality visual perception (Mean Opinion Score, MOS~\citet{mos}), using a scale of 1 to 5 (i.e., 1 = bad, 2 = poor, 3 = adequate, 4 = good, 5 = excellent). After calculating the average score for each criterion, we created radar charts for each US plane to facilitate visual comparison, as shown in Fig. \ref{reader}. Each line on the radar chart represents a different method, and each vertex corresponds to a specific evaluation criterion.

In the reader study, the two experienced radiologists tended to assign lower scores to images with incorrect categories or those that did not align with clinical expectations. This tendency resulted in lower scores for SD, ControlNet, and Uni-ControlNet across all cases, reflecting the limitations of these methods in generating accurate and clinically relevant images. 
Radiologists also assigned lower scores to the Repaint method in the thalamus dataset, attributing this to its lack of control conditions, which often resulted in the synthesis of irrelevant abdominal anatomical structures and category misclassifications.
 
Notably, FetalFlex demonstrated strong competitive performance in image generation. FetalFlex received scores comparable to real images (GT) in the facial dataset, indicating that its generated outputs closely matched clinical standards. FetalFlex even outperformed GT across several key quality metrics in the abdominal dataset. This highlights the model’s potential for improving diagnostic accuracy and enhancing image quality in challenging cases, particularly where real images may suffer from noise or low resolution.
Overall, the radar charts clearly illustrated that FetalFlex’s results align more closely with expert preferences for clinical fetal US image generation, further validating the model’s applicability and advantages across diverse US imaging scenarios.

\subsection{Controllablity Visualization}
The visualization in Fig. \ref{vis} demonstrates the quality and fidelity of the normal images synthesized by FetalFlex, and the construction of single anatomical abnormalities under controlled condition variations. In the first column, the reference anatomical structures were extracted from real US images, whereas the second column shows the synthesized normal samples guided by normal anatomy. These synthesized images closely replicated the spatial relationships between anatomical features, ensuring strong structural consistency with the real images. Additionally, the remaining four columns illustrate FetalFlex’s capability to generate three types of fetal US abnormal samples: 1) simulating structural displacement by altering the relative positions of anatomical structures; 2) simulating underdevelopment or malformation by changing the relative size of one or more organs; and 3) generating malformed samples by constructing missing anatomical structures.

The layout maps showed clear repositioning of anatomical structures in the visualization of simulated structural displacement, accurately reflecting positional changes while maintaining high image fidelity. 
In cases of size variation, the synthesized results indicated that FetalFlex effectively captures proportional changes in specific structures while preserving the normal appearance of others. For samples with structural absence, FetalFlex demonstrated its ability to accurately represent US images with missing anatomical components, even for subtle structural abnormalities. Moreover, the generated images maintained high image feature-label consistency, further validating FetalFlex's utility in producing clinically meaningful training samples for anomaly detection tasks.

FetalFlex is well adapted to the needs of clinical and downstream tasks because abnormal samples of fetal US are highly diverse and difficult to generalize with fixed rules. FetalFlex can flexibly generate data that match the characteristics of multiple abnormalities by precisely controlling each anatomical substructure of the fetus, thus providing the model with greater adaptability and generalizability when dealing with complex and variable abnormality cases. This method improves the diversity of generated data and strengthens the credibility of clinical applications.
Moreover, we considered the significant diversity in multi-center datasets collected by different US devices and radiologists, such as variations in image clarity and grayscale. Our method preserves the original imaging characteristics while generating diverse image samples. 
Additionally, it retains the unique acquisition habits and features associated with different US devices and practitioners in the synthetic dataset, thereby supporting downstream tasks in deep learning model training.

\section{Discussion and Conclusion}
The acquisition of fetal US images presents several challenges, including operational complexity, reliance on technical expertise, and variability among pregnant women, all of which result in inconsistencies in image quality. In addition, the low prevalence of positive diagnostic cases introduces further complexities. Fetal movement and suboptimal positioning exacerbate these difficulties, creating barriers to obtain accurate and reliable imaging. To address these issues, medical image synthesis techniques offer promising solutions to mitigate data scarcity, reduce annotation costs, and protect patient privacy. Recent studies have demonstrated the effectiveness of diffusion models in data augmentation for natural images, although their application in the medical domain remains limited. One major challenge is the lack of a general framework tailored for US images, which complicates domain adaptation and training from scratch.
Furthermore, the limited availability of medical data restricts the models' capacity to capture complex image features, particularly in rare or absent abnormal cases, where reference knowledge is insufficient. Additionally, synthesized images must meet clinicians' visual standards and perform effectively in downstream tasks. Finally, ensuring that the US image generation process is controllable and editable is crucial for practical clinical applications. 

This study introduced FetalFlex, a general and scalable framework capable of generating data from multiple fetal US datasets in prenatal examinations, with the potential for further extension to additional US categories. FetalFlex leverages anatomical structures as prior knowledge and integrates multimodal conditions as control information, enabling the generation of anatomically controlled, realistic, and diverse US images. The proposed method replaces the CLIP text embedding model used in diffusion models with a pre-alignment module, aligning the sparse features of the layout map with the semantically rich US images. Additionally, we incorporate a conditional inpainting strategy in small medical datasets and employ SSIM and MSE loss functions to ensure global and local synthesis quality. Our two-stage sampling strategy further improved the detail of the generated images while maintaining user editability, ensuring the accurate synthesis of subsequent abnormal images. 
We conducted extensive experiments on internal multi-center datasets, demonstrating the ability of the FetalFlex to generate high-quality fetal US images.
Through downstream classification experiments, we demonstrated that FetalFlex's synthetic results, while ensuring consistent class labels, serve as an effective data augmentation tool and have the potential to replace real data, addressing privacy concerns. Moreover, the base deep learning models' ability to recognize positive samples was significantly improved by altering the layout boxes to generate abnormalities. This provides a powerful and flexible means of creating editable synthetic data for multiple potential abnormalities, which is particularly valuable in the fetal US domain, where the incidence of abnormal cases is extremely low. Ultimately, the synthetic results generated by FetalFlex achieved a level of subjective visual perception by clinicians that closely matches that of real US images.

From a medical perspective, fetal US imaging encompasses complex anatomical features and growth parameters, which are critical for assessing fetal development and abnormalities. FetalFlex, with its precise anatomical control, generates high-quality images that capture detailed organ structures and growth information. Notably, FetalFlex can synthesize biologically meaningful abnormal samples from existing normal data, even in the absence of actual abnormal references. This capability facilitates the generation of OOD data from ID data, offering image-level interpretability.
However, our researchers did not manually clean the collected fetal US dataset, resulting in the inclusion of some low-quality images in training and leading to high data heterogeneity. Future research should focus on data cleaning, retaining only high-quality images, to further enhance the quality and visual appearance of the generated results.

However, currently no standardized metrics exist that quantify FetalFlex's controllable generation capabilities. Therefore, validation was conducted through visualization results and anomaly detection experiments.
In the visualizations, we carefully presented the results generated under conditions where individual anatomical structures were altered, as shown in Fig. \ref{vis}. Specifically, we applied image editing techniques, such as translation, rotation, scaling, and removal, to modify the control information for several small anatomical structures, simulating various fetal abnormalities that may occur in the real world. By integrating two-stage sampling strategy, the proposed FetalFlex framework accurately generates US images corresponding to the edited defect layouts without affecting the unedited anatomical structures. In the third row of Fig. \ref{vis999}, we showcase some real cases of cleft palate. The US image displays an interruption or discontinuity in the fetal palate, represented by an anechoic area. Similarly, the FetalFlex-generated cleft palate image exhibits an identical anechoic area in the palate, closely resembling the real case. This result demonstrates FetalFlex's capability to capture fine anatomical details of the fetus and highlights its potential as a universal framework, capable of simulating a wide range of developmental abnormalities or malformations encountered during fetal examination. 

In the visualized results of the general changes in anatomical structure shown in Fig. \ref{vis999}, we guided the global control information to generate images that closely resemble the US representations of real cases of malformation due to limitations imposed by the types of cases in our retrospective dataset. For example, the results synthesized by FetalFlex in the thalamic dataset can simulate the typical characteristics of ventricular enlargement in actual cases of hydrocephalus, specifically manifested as large areas of anechoic or hypoechoic regions, which appear black and represent the presence of fluid within the ventricles. Similarly, the ``double bubble sign'' is a typical feature of duodenal obstruction or stenosis in the abdominal dataset, appearing as two distinct anechoic or hypoechoic cystic structures (black) in the US image. FetalFlex has not encountered real cases; therefore, it can synthesize two black areas merely by altering the layout map but lacks the characteristic proximity of the double bubble observed in real cases. The synthesized results still play a positive role in model training during anomaly detection experiments; nevertheless, further improvements in the visual quality of medical images are required.

In conclusion, this study introduced FetalFlex, an anatomically controllable US image generation model designed to leverage anatomical information for producing high-fidelity, highly controllable fetal US images across multiple planes. 
This model addressed the challenges of limited training data for deep learning models and may aid researchers and clinicians in better understanding fetal US data under various anatomical conditions. FetalFlex integrates multimodal control conditions and employs a two-stage adaptive sampling strategy, progressing from coarse to fine granularity. Additionally, by imposing enhanced anatomical constraints, FetalFlex can simulate OOD abnormal fetal US images, such as variations in anatomical structure size, structural displacement, or structural deficiencies, based solely on ID training samples. Numerous experiments quantitatively and qualitatively proved the effectiveness of FetalFlex. Overall, our study offers a promising approach for expanding normal and abnormal datasets while addressing data privacy concerns. In future work, we aim to develop a generalized framework for fetal US synthesis by incorporating the robust semantic capabilities of large language models into diffusion models.
\\
\section*{CRediT authorship contribution statement}

\textbf{Yaofei Duan}: Writing - original draft, Validation, Methodology, Formal analysis, Conceptualization. \textbf{Tao Tan}: Writing - review \& editing, Supervision. \textbf{Zhiyuan Zhu}: Validation, Visualization. \textbf{Yuhao Huang}: Writing - review \& editing. \textbf{Yuanji Zhang}: Validation, Data curation. \textbf{Rui Gao}: Data curation. \textbf{Patrick Cheong-Iao Pang}: Writing - review \& editing. \textbf{Xinru Gao}: Data curation \& Annotation. \textbf{Guowei Tao}: Data curation \& Annotation. \textbf{Xiang Cong}: Data curation \& Annotation. \textbf{Zhou Li}: Data curation \& Annotation. \textbf{Lianying Liang}: Data curation \& Annotation. \textbf{Guangzhi He}: Data curation \& Annotation. \textbf{Linliang Yin}: Data curation \& Annotation. \textbf{Xuedong Deng}: Data curation \& Annotation. \textbf{Xin Yang}: Writing - review \& editing, Conceptualization, Supervision. \textbf{Dong Ni}: Funding acquisition, Supervision.

\section*{Declaration of competing interest}

The authors declare that they have no known competing financial interests or personal relationships that could have appeared to influence the work reported in this manuscript.

\section*{Data availability}

The authors do not have permission to share data.

\section*{Acknowledgments}
This work was supported by the grant from Science and Technology Development Fund of Macao (0021/2022/AGJ), National Natural Science Foundation of China (12326619, 62101343, 62171290), Science and Technology Planning Project of Guangdong Province (2023A0505020002), Shenzhen-Hong Kong Joint Research Program (SGDX20201103095613036), and Multi-center Clinical Study of Intelligent Prenatal Ultrasound (ChiCTR2300071300).

\bibliographystyle{unsrtnat}
\bibliography{bib}
\end{document}